\documentclass[preprint,12pt]{elsarticle}

\usepackage{amssymb}

\font\mybb=msbm10 at 12pt
\def\bb#1{\hbox{\mybb#1}}
\def\bZ {\bb{Z}}
\def\bR {\bb{R}}

\def\bH {\bb{H}}
\def\bC {\bb{C}}

\def\bX {\bb{X}}
\def\bP {\bb{P}}

\def\bU {\bb{U}}

\def\bW {\bb{W}}
\def\bI {\bb{I}}

\makeatletter \@addtoreset{equation}{section} \makeatother

\def\slashchar#1{\setbox0=\hbox{$#1$}           
   \dimen0=\wd0                                 
   \setbox1=\hbox{/} \dimen1=\wd1               
   \ifdim\dimen0>\dimen1                        
      \rlap{\hbox to \dimen0{\hfil/\hfil}}      
      #1                                        
   \else                                        
      \rlap{\hbox to \dimen1{\hfil$#1$\hfil}}   
      /                                         
   \fi}

\journal{Annals of Physics}

\begin{document}

\begin{frontmatter}



\title{Supertwistors and massive particles}


\author{Luca Mezincescu }

\address{Department of Physics
University of Miami, \\ Coral Gables, FL 33124, USA}
\ead{mezincescu@server.physics.miami.edu}

\author{Alasdair J. Routh and Paul K. Townsend}
\ead{a.j.routh@damtp.cam.ac.uk, p.k.townsend@damtp.cam.ac.uk}

\address{DAMTP, CMS\\
 Wilberforce Road, Cambridge CB3 0WA}


\begin{abstract}
In the  (super)twistor formulation of  massless (super)particle mechanics, the mass-shell constraint is replaced by a ``spin-shell'' constraint from which the spin content can be read off.   We extend this formalism to   massive (super)particles (with $N$-extended spacetime supersymmetry) in three and four space-time dimensions,  explaining how the spin-shell constraints are related to spin, and  we use  it  to prove equivalence of the massive $N=1$ and BPS-saturated $N=2$ superparticle actions.  We also find the supertwistor form of the action for  ``spinning particles''  with $N$-extended worldline  supersymmetry, massless in four dimensions and massive in three dimensions, and we show how this simplifies special features of the $N=2$ case. 
\end{abstract}


\end{frontmatter}


\tableofcontents

\section{Introduction}

The classical geometric action for a massless relativistic point particle in Minkowski spacetime exhibits a conformal invariance that survives in the quantum theory as the maximal symmetry group of  the particle's relativistic wave equation. In  space-time dimensions $D=3,4$ and $6$ (which we abbreviate to 3D, 4D and 6D) the conformal symmetry of the classical action can be made manifest by re-expressing it in terms of twistors (spinors of the conformal group \cite{Penrose:1986ca}) or supertwistors  in the case of superconformal symmetry \cite{Ferber:1977qx,Shirafuji:1983zd,Bengtsson:1987si}. This formulation  has the advantage that covariant quantization leads directly to a manifestly (super)conformal invariant form of the particle's wave equation. In addition,  the usual mass-shell constraint is replaced by a ``spin-shell'' constraint from which the spin-content can be read off; in the case of the superparticle  \cite{Brink:1981nb,Siegel:1983hh}, with manifest space-time supersymmetry, this provides a simple way\footnote{It is simple because it circumvents problems associated with the ``kappa-symmetry'' (a fermionic gauge invariance) of the usual superparticle action; in an an alternative approach,  a twistor-like action provides a geometrical
interpretation of this gauge invariance \cite{Sorokin:1989zi}.} to see that quantization yields a free field supermultiplet.

Spin can also be introduced via (local) worldline supersymmetry \cite{Barducci:1976qu}; the simplest relativistic example being the  massless ``spinning particle'' that yields, upon quantization, the massless Dirac equation \cite{Brink:1976sz}. This action is conformal, but not superconformal invariant; nevertheless, for $D=4$ it too has a supertwistor formulation \cite{Townsend:1991sj}, albeit one in which the Hamiltonian constraints are conformal but not superconformal invariant. There is an $N$-extended generalisation of the 
massless 4D ``spinning particle''  that yields, upon quantization, the free field equation for spin $N/2$  \cite{Gershun:1979fb,Howe:1988ft,Volkov:1989ky}. We show here that this too has a  supertwistor formulation.  The $N=2$ case is special because it allows the addition of a ``worldline Chern-Simons term'' which leads, upon quantisation,  to a description of spin-zero by an antisymmetric tensor field \cite{Howe:1989vn}; the spin-zero content of this model becomes manifest in its twistor formulation. 

More surprisingly, there is a similar  (super)twistor formulation for {\it massive} point particles in $D=3,4$, although it requires
a pair of (super)twistors  \cite{Hughston:1981zc}.  This can be partly understood from the fact that a massive particle in $D=4(3)$ dimensions can be viewed as a dimensionally reduced massless particle in $D=6(4)$, where dimensional reduction, in this context,  amounts to  a constraint that specifies the higher-dimensional components of the momentum.  This construction has been applied in \cite{deAzcarraga:2008ik,Fedoruk:2013sna} and it leads directly to a bi-twistor formulation of massive particles\footnote{There are various other ways to understand why a bi-twistor formulation of massive particles is possible, e.g. \cite{Fedoruk:2003td} or the tractor formalism \cite{Gover:2008sw}.}. In particular, 
the supertwistor formulation of the massless 6D superparticle can be dimensionally reduced to yield the bi-supertwistor formulation of  the 4D  ``BPS superparticle''  with $N=2$ supersymmetry \cite{deAzcarraga:2008ik}.

The generic $N=2$ 4D superparticle action has two mass parameters, the (positive) mass $m$ appearing in the mass-shell constraint and a coefficient $q$ of a Wess-Zumino (WZ) mass term \cite{Azcarraga:1982dw}. Unitarity of the quantum theory (absence of negative norm states) imposes the bound $m\ge |q|$, which we shall refer to  as the ``BPS bound''.  
By ``BPS superparticle'' we mean one for which the BPS  bound is saturated, i.e.  $m=|q|$. One purpose of this paper is to show, by example,  that the (super)twistor formulation of massive 4D particle mechanics is not restricted to the BPS-saturated case\footnote{This has also been done in \cite{Fedoruk:2005ks}, and by similar means, although it is at present unclear to us what the relation is to the results reported here.}.  In particular, we construct the supertwistor form of the massive $N=1$ 4D superparticle action; 
this turns out to be  {\it identical}  to the supertwistor form of the $N=2$ BPS superparticle. 

Because the twistor form of the action for a massive 4D particle can be simply obtained from that of the massless 6D particle, we review this construction here. At the same time we present some new results on the 6D case. As shown in \cite{Bengtsson:1987si} the spin-shell constraint functions of the twistor form of the massless 6D particle generate an ``internal'' $SU(2)$ gauge invariance, which has no obvious connection to ``spin''.  Here we show that these constraint functions are 6D analogs of  4D helicity.  
Furthermore, we show that the Poincar\'e Casimir of the massive 4D particle found from the square of the Pauli-Lubanski spin-vector is proportional, when expressed in twistor variables, to the quadratic Casimir of   the ``internal''  $SU(2)$, thereby justifying the terminology ``spin-shell''  for the constraints of this model. 

In the 3D case, it has been appreciated for some time that a bi-twistor  formulation of massive particle mechanics is possible \cite{Sorokin:1992sy,Gorbunov:1996ed}, but in most formulations the twistor variables have been additional to the standard ones;  for purposes of comparison we present a brief review of the spinless point particle action of Sorokin and Volkov \cite{Sorokin:1992sy}. A (super)twistor formulation of massive 3D particle mechanics in which the (super)twistor variables are {\it alternatives} to the usual phase-space  was discussed briefly in \cite{Fedoruk:2013sna}, and here we further develop this formalism. As  in the 4D case, we find that the $N=1$ massive superparticle action is identical to the $N=2$ BPS superparticle action, in agreement with  Gorbunov et al. \cite{Gorbunov:1996ed}  (see also \cite{Gorbunov:1998id}).   We also obtain the supertwistor form of the $N$-extended massive 3D ``spinning particle''.  

A special feature of 3D massive particles  is that we may add to the action  the parity-violating ``Lorentz-Wess-Zumino'' (LWZ) term \cite{Schonfeld:1980kb}. The effect, in the quantum theory,  is to  shift  all particle helicities by the coefficient of this  term, which can be any real number\footnote{By ``helicity''  we mean, in the 3D context, the value of the 3D Pauli-Lubanski scalar divided by the mass; we reserve  the  term ``spin'' for  its absolute value.}. As we show here, this too can be seen very simply in our  supertwistor formulation (as is also true of the formalism of  \cite{Gorbunov:1996ed}). 

We begin with a  summary of our notation and conventions and a review of the  construction of a twistor form of the action for a massless particle in dimensions $D=3,4$ and $6$. We then go on to consider massive 3D particles and various possible ways in which spin can be introduced, explaining  in each case how the action can be written in Hamiltonian form with a phase space parametrized by 3D (super)twistors.  We then proceed to consider various  4D cases for which a (super)twistor formulation is possible, and to apply the results in the way described. 
We conclude with a summary and a discussion of some of the finer points. 

\section{Preliminaries}\label{sec:prelim}
In what follows, we assume a Minkowski spacetime of dimension $D=3,4$ or $6$. In these dimensions the spin group is, respectively,  $Sl(2;\bR)$, $Sl(2;\bC)$ and 
$Sl(2;\bH)\cong SU^*(4)$.  The metric signature is ``mostly plus''.

\subsection{Spinor conventions}
We begin with a summary of our spinor conventions. 

\begin{itemize}

\item We take the coordinates of the $N$-extended 3D superspace to be $\left\{x^\mu,p_\mu; \theta_a\right\}$ ($a=1,\dots,N$) where $x^\mu$ and $p_\mu$ are $3$-vectors and 
$\theta_a$ (spinor indices suppressed) are $N$ anticommuting $2$-component Majorana spinors.  
We denote by $\gamma$ the 3D Dirac matrices. A convenient real representation is 
\begin{equation}
\gamma^0= i\sigma_2\, , \qquad \gamma^1= \sigma_1\, , \qquad \gamma^2 = \sigma_3\,  \qquad \left(\Rightarrow \quad \gamma^0\gamma^1\gamma^2=1\right). 
\end{equation}
We may choose the 3D charge conjugation matrix to be $\gamma^0$, in which case a Majorana spinor $u$  is real and its conjugate is $\bar u = u^T \gamma^0$. 
Any commuting 3D Majorana spinor satisfies the identity
\begin{equation}\label{3Did}
\left(\bar u \gamma u\right) \cdot \gamma u \equiv 0\, . 
\end{equation}

We also explain here our  3D spinor index conventions. We let $u^\alpha$ ($\alpha=1,2)$ denote the components of a 3D spinor, and $\bar u_\alpha$ the components of its conjugate; 
if $u$ is Majorana then 
\begin{equation}
\bar u_\alpha= u^\beta \varepsilon_{\beta\alpha}\, , \qquad u^\alpha = \varepsilon^{\alpha\beta} \bar u_\beta\, , 
\end{equation}
where $\varepsilon^{\alpha\beta}$ is numerically equal to $\varepsilon_{\alpha\beta}$ and hence satisfies
\begin{equation}\label{prodeps}
\varepsilon^{\alpha\beta}\varepsilon_{\beta\gamma}= - \delta^\alpha_\gamma\, . 
\end{equation}
This  relation can be interpreted as the raising of an  index by  $\varepsilon^{\alpha\beta}$ or the lowering of an index by $\varepsilon_{\beta\gamma}$ provided that one 
uses  the  convention for which 
\begin{equation}
\varepsilon_\alpha{}^\beta = \delta_\alpha^\beta\, , \qquad \varepsilon^\alpha{}_\beta = - \delta^\alpha_\beta\, .  
\end{equation}
As a consistency check, we note that (\ref{prodeps}) implies that  $\varepsilon^{\alpha\gamma}\varepsilon^{\beta\delta}\varepsilon_{\gamma\delta} = \varepsilon^{\alpha\beta}$, 
which is what we expect from the index raising interpretation. 

A feature of these conventions  is that we may write any 4D Majorana spinor $U$ as 
\begin{equation}
U= \left( \begin{array}{c} u\\ v \end{array} \right)\, , 
\end{equation}
where $(u,v)$ are a pair of 3D Majorana spinors. Then 
\begin{equation}\label{P3}
\bar U \Gamma_3 U =- 2\bar u v\, , 
\end{equation}
and 
\begin{equation}\label{3DP}
\bar U \Gamma^\mu U = \bar u \gamma^\mu u + \bar v\gamma^\mu v \qquad (\mu=0,1,2). 
\end{equation}
Given another Majorana spinor 
\begin{equation}
W =  \left( \begin{array}{c} z\\ w \end{array} \right)\, , 
\end{equation}
we have 
\begin{equation}\label{kinetic4}
\dot{\bar U} W = \dot{\bar u} w +\dot{\bar v} z\, , \qquad \bar U \gamma_5 W  = \bar u z - \bar v w\, . 
\end{equation}

\item We take the coordinates of the $N$-extended 4D superspace to be $\left\{X^m,P_m; \Theta_a\right\}$ ($a=1,\dots,N$) where $X^m$ and $P_m$ are $4$-vectors and $\Theta_a$ (spinor indices suppressed) are $N$ anticommuting $4$-component Majorana spinors.  We denote by $\Gamma$ the 4D Dirac matrices. A convenient real representation is 
\begin{equation}
\Gamma= \left(\gamma \otimes \sigma_1, \bI \otimes \sigma_3\right) \, , 
\end{equation}
where $\gamma$ are the $2\times 2$ 3D Dirac matrices. This choice implies that 
\begin{equation}
\gamma_5 \equiv \Gamma^0\Gamma^1\Gamma^2\Gamma^3 = -\bI \otimes i\sigma_2\, . 
\end{equation}
We may choose the 4D charge conjugation matrix $C$ to be $\Gamma^0$, in which case a Majorana spinor $U$ is real and its conjugate is $\bar U = U^T \Gamma^0$. Any commuting 4D Majorana spinor $U$ satisfies the identity 
\begin{equation}\label{4Did}
\left(\bar U \Gamma U\right) \cdot \Gamma U \equiv 0\, . 
\end{equation}

\item For the 6D particle, we use an $SU^*(4)$ spinor notation  \cite{Koller:1982cs,Howe:1983fr}  in which spinors are 4-component $SU(2)$ doublets and 
vectors are anti-symmetric bi-spinors. In this notation,   the phase space coordinates are $(\bX^{\alpha\beta},\bP_{\alpha\beta})$ ($\alpha,\beta=1,2,3,4$). 
We do not need to specify 6D superspace notation  here as we will not discuss the  6D superparticle, which was dealt with in  \cite{Bengtsson:1987si,deAzcarraga:2008ik}.

We can decompose a 6D spinor doublet $\bU^i$  into two 4D Majorana spinors $(U,V)$ as follows:
\begin{equation}
\mathbb{U}^1 = \frac{1}{\sqrt{2}}\left(U + iV\right) \, , \qquad \bU^2 = -\frac{1}{\sqrt{2}} \gamma_5\left( U-iV \right)\,.
\end{equation}
A spinor doublet $\bW_i$ of the opposite chirality and a lowered $SU(2)$ index decomposes into the conjugates of two 4D  Majorana spinors $(W,Z)$ as follows:
\begin{equation}
\mathbb{W}_1 = -\frac{1}{\sqrt{2}}\left(\bar{W} - i\bar{Z} \right) \, , \qquad  \bW_2= -\frac{1}{\sqrt{2}}\left( \bar{W}+ i\bar{Z}\right)\gamma_5 \,.
\end{equation}
This leads, for example, to
\begin{equation}
\dot{\mathbb{U}}_\alpha^i \mathbb{W}^\alpha_i = \dot{\bar{U}}W + \dot{\bar{V}}Z \,.
\end{equation}
Similarly,
\begin{eqnarray}\label{6Dto4D}
-i\mathbb{W}_i^\alpha \sigma_1{}^i{}_j\mathbb{U}^j_\alpha &=& \bar{V}\gamma_5 W + \bar{U}\gamma_5 Z  \, , \nonumber \\
-i\mathbb{W}_i \sigma_2{}^i{}_j\mathbb{U}^j &=& \bar{U}\gamma_5 W - \bar{V}\gamma_5 Z \, ,  \nonumber \\
-i\mathbb{W}_i \sigma_3{}^i{}_j\mathbb{U}^j &=& \bar{V}W - \bar{U}Z  \, . 
\nonumber \\
\end{eqnarray}

\end{itemize}

\subsection{Twistor action for massless particles}

We now summarize  the twistor formulation of the massless spin-zero particle  in 3D, 4D and  6D.

\begin{itemize}

\item  The action for a massless 3D particle is 
\begin{equation}\label{3Dpoint}
S= \int \! dt \left\{ \dot x \cdot p -  \frac{1}{2} e\, p^2  \right\} \, , 
\end{equation}
where $e$ is a Lagrange multiplier for the mass-shell constraint. As a consequence of the identity (\ref{3Did}), this constraint has the solution
\begin{equation}\label{solveP}
p= -\frac{1}{2} \bar u \gamma u\, , 
\end{equation}
where $u$ is a Majorana spinor. The sign is chosen such that $p^0>0$. Substitution  yields\footnote{The precise form of  the total time derivative  will not be needed so it suffices to  indicate its presence  by empty parentheses. We use this shorthand, when appropriate, throughout the paper.}
\begin{equation}
\dot x\cdot p = \dot{\bar u} w + \frac{d}{dt} \left(\right)\, , \qquad w= \slashchar{x} u \, . 
\end{equation}
Observe that $w$ is unchanged, as a consequence of (\ref{3Did}), by the infinitesimal gauge transformation $x\to x + \alpha(t) p$, with parameter $\alpha(t)$,  which is generated by the mass-shell constraint function. Because of this gauge transformation only two components of $x$ are physical, and we may trade these for $w$, to arrive at the action
$S= \int \! dt \, \dot{\bar u}w$. The pair of spinors $(u,w)$ are components of a 3D twistor, i.e. a spinor of the 3D conformal group $SO(2,3)$ or, equivalently, a real $4$-vector of its double cover $Sp(4;\bR)$. The four real components of this twistor parametrize the physical phase space of the massless 3D particle.

\item The action for a massless 4D point particle is 
\begin{equation}\label{4Dpoint}
S= \int \! dt \left\{ \dot X \cdot P -  \frac{1}{2} e P^2  \right\} \, .
\end{equation}
As a consequence of the identity (\ref{4Did}), the mass-shell constraint has the solution
\begin{equation}\label{solveP4D}
P= -\frac{1}{2} \bar U \Gamma U\, , 
\end{equation}
where $U$ is a Majorana spinor. As P is unchanged by the infinitesimal gauge transformation 
\begin{equation}
U \to U + \beta(t)\gamma_5 U\, , 
\end{equation}
for parameter $\beta(t)$,  we should expect a corresponding first-class constraint in the twistor form of the action. 
Substitution for $P$ yields
\begin{equation}
\dot X\cdot P = \dot{\bar U} W + \frac{d}{dt} \left(\right)\, , \qquad W= \slashchar{X} U\, . 
\end{equation}
As in the 3D case we  aim to promote $W$ to the status of an independent variable, but it follows from its definition that 
\begin{equation}
\bar U \gamma_5 W \equiv 0\, .
\end{equation}
This must be added as constraint to the action in terms of the new canonical spinor variables $(U,W)$; this is the constraint expected from the
$U(1)$ gauge invariance introduced by the solution for $P$.  The result is the action
\begin{equation}\label{4Dtwistorpart}
S= \int \! dt \left\{ \dot{\bar U} W  - \frac{1}{2}s\, \bar U \gamma_5 W \right\}\, , 
\end{equation}
where $s$ is a Lagrange multiplier for the ``spin-shell'' constraint  $\bar U \gamma_5 W =0$.  The Majorana spinors $(U,W)$  are the components of a 4D twistor, i.e. a spinor of the 4D conformal group $SO(2,4)$ or, equivalently, a complex $4$-vector  of  its double cover $SU(2,2)$. 

\item 
In $SU^*(4)$ spinor notation the action for a massless particle in 6D is 
\begin{equation}
S= \int\! dt\left\{ \dot {\bX}^{\alpha\beta} \bP_{\alpha\beta} - \frac{1}{2} e\,  \bP^2 \right\} \, , 
\qquad {\bP}^2 = \frac{1}{2}
\varepsilon^{\alpha\beta\gamma\delta} \bP_{\alpha\beta} \bP_{\gamma\delta} \, . 
\end{equation}
We can solve the mass-shell constraint, in terms of an $SU(2)$ doublet $\bU^i$ $(i=1,2)$ of  $SU^*(4)$ spinors, by writing
\begin{equation}\label{Psq}
\bP_{\alpha\beta} = \frac{1}{2} \bU_\alpha^i \bU_\beta^j \, \varepsilon_{ji} \, .
\end{equation}
Notice that this solution of the mass-shell constraint is invariant under local $SU(2)$ transformations of $\bU^i$.  Substitution for $\bP$ also gives
\begin{equation}\label{defW6}
\dot{\bX} \cdot \bP  = \dot{\bU}_\alpha^i  \bW^\alpha_i + \frac{d}{dt} \left(\right) \, , \qquad \bW^\alpha_i= X^{\alpha\beta}\bU_\beta^j \varepsilon_{ji}\, . 
\end{equation}

Let us define 
\begin{equation}\label{6DLam}
\Lambda^i{}_j = \left(\bU\bW\right)^i{}_j - \frac{1}{2} \delta^i_j (\bU\bW) \, , 
\end{equation}
where
\begin{equation}
 (\bU\bW)^i{}_j = \bU_\alpha^i \bW^\alpha_j \, , \quad (\bU\bW) = (\bU\bW)^i{}_i\, .
\end{equation}
Observe that $\Lambda^i{}_j$ is traceless, which is equivalent to symmetry, in its $SU(2)$ indices, of $\Lambda^{ij} = \varepsilon^{jk} \Lambda^i{}_k$. 
We use the same conventions to raise and lower $SU(2)$ indices as those explained earlier for raising and lowering $Sl(2;\bR)$ spinor indices. 

Given the definition of $\bW_i$ in (\ref{defW6}), we have $\Lambda^i{}_j\equiv0$, so this becomes a constraint when $\bW$ is considered as an independent variable. 
This  gives us  the following twistor form of the action for a massless 6D particle:
\begin{equation}\label{6Dact}
S= \int\! dt \left\{ \bU_\alpha^i \dot {\bW}^\alpha_i  - s_{ij} \Lambda^{ij} \right\}\, . 
\end{equation}
The $SU(2)$ triplet $\Lambda^{ij}$ generates the local $SU(2)$ gauge transformations, via the canonical Poisson bracket relations
\begin{equation}\label{PB6D}
\left\{ \bU_\alpha^i , \bW^\beta_j \right\}_{PB} = \delta_\alpha^\beta \delta^i_j \, . 
\end{equation}

The Poincar\'e Noether charges in $SU^*(4)$ spinor notation are
\begin{equation}\label{Noethertwist}
{\cal J}_\alpha{}^\beta = \bU_\alpha^i \bW^\beta_i - \frac{1}{4} \delta_\alpha^\beta \left(\bU\bW\right)\, , 
\qquad {\cal P}_{\alpha\beta} = \bP_{\alpha\beta} \equiv \frac{1}{2} \bU_\alpha^i \bU_\beta^j \varepsilon_{ji}\, .
\end{equation}

\end{itemize}

\subsection{Spin-shell constraints and generalized helicities}\label{subsec:genhel}

In the 4D case the spin-shell constraint function generates an ``internal'' $U(1)$ gauge invariance, and the constraint sets to zero the $U(1)$ charge. 
In the 6D case the spin-shell constraint functions generate an ``internal'' $SU(2)$ gauge invariance, and the constraints set to zero an $SU(2)$ triplet  charge. 
This raises the question of how these $U(1)$ or $SU(2)$  charges are related to the particle's spin.  The answer is known in the 4D case, but the issue has not  yet 
been addressed, as far as we are aware, in the 6D case. We shall consider the 4D and 6D cases in turn.

\begin{itemize}

\item For a 4D particle, the Pauli-Lubanski spin-vector is
\begin{equation}
\Sigma^m = \varepsilon^{mnpq} {\cal J}_{np} {\cal P}_q\, ,  
\end{equation}
where $({\cal J},{\cal P})$ are the particle's Poincar\'e Noether charges.  For a massive particle, $\Sigma^2$ is a Casimir and its value determines
the particle's spin. All Poincar\'e Casimirs are zero for a massless  particle,  but in this case $\Sigma$ equals the particle's 
helicity times its 4-momentum, and the helicity determines the particle's  spin. Using the twistor form of the Poincar\'e Noether charges for a massless 4D particle, 
one finds that 
\begin{equation}
\Sigma^m= \Lambda P^m\, , \qquad \Lambda =  -\frac{1}{2}\left( \bar U \gamma_5 W\right) \, , 
\end{equation}
This shows that the $U(1)$ constraint function  $\Lambda$ of the twistor form of the action (\ref{4Dtwistorpart}) is the particle's helicity. The spin-shell constraint of this action 
tells us that this is zero, as expected for a particle of zero spin.

\item  In the 6D case we have the following 3-form generalization of the Pauli-Lubanski 4-vector
\begin{equation}
\Sigma^{MNP} = \varepsilon^{MNPQRS } {\cal J}_{QR} {\cal P}_S\, . 
\end{equation}
This can be decomposed into a self-dual and anti-self-dual 3-form. In $SU^*(4)$ notation these are the symmetric bispinors
\begin{eqnarray}
\Sigma^{(+)}_{\alpha\beta} &\equiv& {\cal J}_{(\alpha}{}^\gamma {\cal P}_{\beta)\gamma} = \frac{1}{2} \Lambda_{ij} \bU_\alpha^i \bU_\beta^j \, , \nonumber \\
\Sigma_{(-)}^{\alpha\beta} &\equiv& {\cal J}_\gamma{}^{(\alpha}{\cal P}^{\beta)\gamma} =0\, , 
\end{eqnarray}
where $\Lambda_{ij}$ is the $SU(2)$ triplet of constraint functions given in (\ref{6DLam}), and the second equality, in each case, is found upon using (\ref{Noethertwist}). 

The scalar  found by contraction of $\Sigma^{(+)}$ with $\Sigma_{(-)}$ is obviously zero, 
as expected since all Poincar\'e Casimirs are zero for a massless particle, but we see that not only  do $\Lambda_{ij}$ generate the local 
$SU(2)$ invariance but also that they generalize to 6D the notion of  4D helicity. They are not  Casimirs because they are not expressible in terms of the Poincar\'e 
charges but their Poisson brackets with these charges are zero.  For example, the 
canonical Poisson bracket relations (\ref{PB6D}) imply that
\begin{equation}
\left\{ \left(\bU\bW\right)^i{}_j, \bP_{\alpha\beta}\right\}_{PB} = - \delta^i_j \, \bP_{\alpha\beta}\, , 
\end{equation}
and from this  it follows that $\left\{\Lambda_{ij},{\cal P}_M\right\}_{PB}=0$ . It follows that  $SU(2)$ irreps will correspond to unitary Poincar\'e irreps.  

There is a further Casimir in 6D, obtained by squaring the vector:
\begin{equation}
\Xi^M = \varepsilon^{MNPQRS} {\cal J}_{NP} {\cal J}_{QR} {\cal P}_S \, . 
\end{equation}
In $SU^*(4)$ spinor notation we have (ignoring an overall factor)
\begin{equation}\label{J}
\Xi_{\alpha\gamma} \ =\ {\cal J}_\alpha{}^\beta {\cal J}_\gamma{}^\delta {\cal P}_{\beta\delta} - \frac{1}{4}\, {\cal J}_\beta{}^\delta {\cal J}_\delta{}^\beta\, {\cal P}_{\alpha\gamma}\, . 
\end{equation}
The relative factor can be determined by the requirement that $\Sigma$ have zero Poisson bracket with ${\cal P}$.
Using the expressions (\ref{Noethertwist}) we find that 
\begin{equation}
\Xi^M= -\frac{3}{4}\left( \Lambda^k{}_i \Lambda^i{}_k \right)\,  \bP^M\, , 
\end{equation}
which shows that the quadratic $SU(2)$ Casimir is another 6D analog  of 4D helicity.  

\end{itemize}

\section{Massive 3D (super)particle}\label{sec:3D}

 By setting $P_3=m$ in the action (\ref{4Dpoint}) we get the action for a massive 3D particle. We will now investigate where this procedure leads  if we start from the twistor
form of the 4D massless particle action.  From (\ref{3DP}) we see that  (\ref{solveP}) becomes
\begin{equation}\label{particlep}
p =  - \frac{1}{2}\left(\bar u \gamma u + \bar v \gamma v\right)\, , 
\end{equation}
which indeed solves the 3D mass-shell constraint $p^2+m^2=0$ as a consequence of  (\ref{3Did}) and the further identity
\begin{equation}
\left(\bar u \gamma u \right)\cdot \left(\bar v \gamma v \right) \equiv -2 \left(\bar u v\right)^2\, . 
\end{equation}
The solution is the general one with $p^0>0$. From (\ref{P3}) we see that the 3D Majorana spinors $(u,v)$ are constrained to satisfy
\begin{equation}
\bar u v =m \, . 
\end{equation}
Using (\ref{kinetic4}) we then get the  action 
\begin{equation}\label{lag3D}
S= \int \! dt \left\{ \dot{\bar u} w + \dot{\bar v}  z  - \ell \left(\bar u v-m\right)- s \Lambda\right\}\, ,   
\end{equation}
where $\ell$ is a new Lagrange multiplier for the constraint $\bar u v=m$, and 
\begin{equation}
\Lambda = \frac{1}{2} \left( \bar u z - \bar v w \right) \, . 
\end{equation}
The spinor pairs $(u,w)$ and $(v,z)$ are the components of two 3D twistors.  However, the constraint $\bar uv =m$  breaks  conformal invariance to 3D Lorentz invariance.

We may read off from the action (\ref{lag3D}) that the non-zero Poisson bracket relations of the canonical variables are 
\begin{equation}
\left\{\bar u_\alpha, w^\beta\right\}_{PB} = \delta_\alpha^\beta \, , \qquad 
\left\{\bar v_\alpha,z^\beta\right\}_{PB} = \delta_\alpha^\beta \, . 
\end{equation}
Given these Poisson brackets, it follows that the two constraint functions have zero Poisson bracket and hence that they are both ``first-class'' in Dirac's terminology. 
This means that the constraint functions generate gauge invariances, implying a physical phase space of dimension $8-2\times 2=4$, as expected. The gauge transformation 
generated by the constraint function $\bar u v-m$   is  on-shell equivalent  to a reparametrization of the time coordinate $t$. The spin-shell constraint function  $\Lambda$ generates  
a  $U(1)$ gauge transformation that shifts the phase of the complex 3D spinors $u+iv$ and $z+iw$.

Let us now check that the action (\ref{lag3D}) describes a particle of zero spin. To do this we need to find the 
Noether charges resulting from 3D Poincar\'e invariance. The spin-shell constraint can be solved by setting 
\begin{equation}
w = \slashchar{x} u \, , \qquad z= \slashchar{x} v\, , 
\end{equation}
and substitution shows that $p$, as given by (\ref{particlep}) is the momentum conjugate to $x$; this takes us back to the action in terms of the phase space coordinates $(x,p)$.
For present purposes we observe that the space-time translation $x\to x+ a$ is equivalent to 
\begin{equation}\label{wz}
w \to w + a \cdot \gamma u \, , \qquad  z \to  z + a\cdot \gamma v\, . 
\end{equation}
This is indeed a symmetry of the action (\ref{lag3D}), and the corresponding  Noether charge is
\begin{equation}
{\cal P} = -\frac{1}{2} \left(\bar u \gamma u + \bar v \gamma v\right)\, , 
\end{equation} 
as expected.  The Lorentz transformation of $u$ is $\delta u = \frac{1}{2} \slashchar{\omega} u$, where $\omega$ is a $3$-vector parameter, and similarly for the other 
canonical spinor variables.  The corresponding $3$-vector  Noether charge is
\begin{equation}
{\cal J}=  \frac{1}{2}\left(\bar u \gamma  w + \bar v \gamma z\right)\, . 
\end{equation}
Using the identities
\begin{equation}\label{uvids}
\left(\bar u\gamma u\right) \cdot\gamma v = 2 \left(\bar u v\right) u \, , \qquad 
\left(\bar v\gamma v\right) \cdot\gamma u = -2 \left(\bar u v\right) v\, , 
\end{equation}
we deduce that 
\begin{equation}
{\cal P} \cdot {\cal J} = \frac{1}{2}\left( \bar u v \right) \left( \bar u z - \bar v w\right)\, . 
\end{equation}
Then, using the constraint $\bar u v=m$, we  find that the 3D helicity is 
\begin{equation}
 m^{-1} {\cal P} \cdot {\cal J} =  \frac{1}{2} \left( \bar u  z - \bar v w\right) \equiv \Lambda\, , 
\end{equation}
but this is zero as a consequence of the spin-shell constraint.

\subsection{Arbitrary spin case} 

The action for a 3D particle of mass $m$ and helicity $\lambda$ is \cite{Schonfeld:1980kb}
\begin{equation}
S= \int\! dt \left\{ \dot x \cdot p - \frac{1}{2}e \left(p^2 +m^2\right) - \lambda L_{LWZ}\right\}\, , 
\end{equation}
where $L_{LWZ}$ is the Lorentz-Wess-Zumino (LWZ) term. This is the integral of the 1-form $\Omega$, defined locally by 
\begin{equation}
d\Omega = \frac{1}{2\left(-p^2\right)^{\frac{3}{2}}} \varepsilon^{\mu\nu\rho} p_\mu\,  dp_\nu \wedge dp_\nu\, . 
\end{equation}
Under parity we have 
\begin{equation}
x_2 \to - x_2\, , \qquad p_2 \to -p_2\, . 
\end{equation}
which implies that parity is broken only by the LWZ term. 

We now solve the mass-shell constraint as before. Using (\ref{particlep}) we find that 
\begin{equation}\label{LWZuv}
d\Omega =    \frac{1}{m} \left(d\bar u\wedge du +  d\bar v \wedge dv\right)\, . 
\end{equation}
We now have 
\begin{equation} 
L_{LWZ} = -\frac{1}{m} \left(\dot{\bar u} u + \dot{\bar v} v\right)\, . 
\end{equation}
Adding this term leads to a modification of the expressions (\ref{wz}) for the spinors canonically conjugate to $(u,v)$, and the spin-shell constraint must now be 
rewritten in terms of these new variables. After using the constraint $\bar u v=m$ to simplify the result, we arrive at the action 
\begin{equation}\label{neweract}
S= \int \! dt \left\{  \dot{\bar u}  w  +  \dot{\bar v}  z  - \ell\left(\bar u v-m\right) - s \left(\Lambda-\lambda\right) \right\}\, .
\end{equation}
The spin-shell constraint of (\ref{neweract}) now imposes the condition $\Lambda = \lambda$, 
confirming that the particle has helicity $\lambda$, and hence spin $|\lambda|$. 

\subsection{Quantization}

For quantization purposes, it is convenient to rewrite  the action (\ref{neweract}) as 
\begin{equation}
S= \int \! dt \left\{ \dot{\bar \rho} \omega  - \bar\omega \dot\rho  +i \ell \left(\bar \rho\rho -  im\right) - s\left(\Lambda -\lambda\right) \right\}\, , 
\end{equation}
where the complex 3D spinors $(\rho,\omega)$ (components of a complex 3D twistor) and their conjugates are 
\begin{eqnarray}\label{compspin}
\rho&=& \frac{1}{\sqrt{2}}\left(u+iv\right)\, , \qquad \bar\rho= \frac{1}{\sqrt{2}}\left(\bar u -i \bar v\right)\, , \nonumber \\
\omega &=& \frac{1}{\sqrt{2}}\left(w+iz\right)\, , \qquad \bar\omega= \frac{1}{\sqrt{2}}\left(\bar w -i \bar z\right)\, , 
\end{eqnarray}
and
\begin{equation}
\Lambda= -\frac{i}{2} \left(\bar\rho\omega+ \bar\omega\rho\right)\, . 
\end{equation}
Upon quantization, the Poisson bracket relations that follow from this action  are replaced by the canonical commutation relations
\begin{equation}
\left[ \bar\rho_\alpha, \omega^\beta\right] =i \delta_\alpha^\beta\, , \qquad \left[\bar\omega_\alpha,\rho^\beta\right] = i\delta_\alpha^\beta\, , 
\end{equation}
which may be realized on wavefunctions $\Psi(\rho, \bar\rho)$ by setting 
\begin{equation}
\omega^\alpha \to -i\frac{\partial}{\partial\bar\rho_\alpha}\, , \qquad 
\bar\omega_\alpha \to i \frac{\partial}{\partial\rho^\alpha} \, .
\end{equation}
The spin-shell constraint then becomes the physical state condition 
\begin{equation}
\left(\rho^\alpha \frac{\partial}{\partial\rho^\alpha}- \bar\rho_\alpha \frac{\partial}{\partial \bar\rho_\alpha} \right)\Psi = 2\lambda \Psi\, . 
\end{equation}
This must be supplemented by the configuration space constraint $\bar\rho\rho= im$, which implies that
\begin{equation}
2\rho^\alpha \bar\rho_\beta=  \left(p\cdot \gamma\right)^\alpha{}_\beta + im \delta^\alpha_\beta \, , 
\end{equation}
where $p$ is the 3-momentum as given by  (\ref{particlep}). 

If we assume that $\Psi(\varphi,\bar\varphi)$ has a power series expansion then we require $2\lambda \in \bZ$; in this case we may assume that $2\lambda=2s$, a positive integer, and then
\begin{equation}
\Psi = \psi_{\alpha_1\cdots \alpha_{2s}}(p) \rho^{\alpha_1} \cdots \rho^{\alpha_{2s}}\, , 
\end{equation}
for arbitrary multi-spinor coefficient function $\psi$. 
This is the momentum space solution of the standard 3D wave-equation for a particle of spin $s$.  In the case that $2s \notin \bZ$ the solution for $\Psi$ is not of power series 
form \cite{Gorbunov:1996ed}.

\subsection{Comparison with Sorokin-Volkov  action}

The action of Sorokin and Volkov \cite{Sorokin:1992sy} for a 3D massive particle of zero spin  is, in our notation, 
\begin{equation}\label{SV}
S_{SV}= \int\! dt \left\{ \dot x \cdot p + e\left[\frac{1}{2} p\cdot \left(\bar u \gamma u + \bar v \gamma v \right) - m\bar uv\right]\right\}\, . 
\end{equation}
By varying with respect to $x$ and $p$ we get 
\begin{equation}\label{xpeqs}
\dot p=0\, , \qquad e^{-1} \dot x = - \frac{1}{2}\left(\bar u\gamma u +\bar v\gamma v\right)\, . 
\end{equation}
By varying with respect to $u$ and $v$ we get the equations
\begin{equation}\label{uveqs}
\slashchar{p} \, u = mv\, , \qquad \slashchar{p} \,   v= -mu\, . 
\end{equation}
Clearly, these equations are solved by $u=v=0$, but we  shall exclude this solution. It follows from this assumption that 
\begin{equation}
\bar u v\ne0\, . 
\end{equation}
To see this, suppose that $\bar u v=0$; then equations (\ref{uveqs}) imply that $p$ is orthogonal to both $\bar u\gamma u$ and $\bar v\gamma v$, which means that $p\propto \bar u \gamma v$,  which is null when $\bar u v=0$ since 
\begin{equation}
\left(\bar u \gamma v\right)^2 \equiv \left(\bar u v\right)^2\, . 
\end{equation}
However, equations (\ref{uveqs}) also imply the mass-shell constraint 
\begin{equation}
p^2+m^2 =0\, , 
\end{equation}
from which it follows that $p$ is timelike, and hence that $\bar u v\ne0$. 

The equations (\ref{uveqs}) also imply that 
\begin{equation}
p\cdot \left(\bar u\gamma u + \bar v\gamma v\right) = 2m\, \bar u v\, , 
\end{equation}
which is what we get by varying with respect to $e$.  This is a consequence of a scaling gauge invariance; if $u$ and $v$ are assigned a unit scaling weight then 
$e$ has scaling weight $-2$.  We may fix this gauge invariance by the gauge choice 
\begin{equation}
\bar u v=m\, . 
\end{equation}

So far, we have not found any relation between $p$ and $\dot x$. Let us now define a new 3-vector variable $q$ by 
\begin{equation}
p= - \frac{m}{2\bar u v}\left(\bar u\gamma u + \bar v\gamma v\right) + q\, . 
\end{equation}
 If we substitute for $p$ in (\ref{uveqs}) and use the identities (\ref{uvids}) then we deduce that $\slashchar{q}u=\slashchar{q}v=0$. This implies that $q$ is proportional to both 
$\bar u \gamma u$ and $\bar v\gamma v$, which implies {\it either} that $q=0$ {\it or} that $u\propto v$. But the equations (\ref{uveqs}) do not allow $u\propto v$ for non-zero $m$, so $q=0$, and hence
\begin{equation}\label{puv}
p= - \frac{m}{2\bar u v}\left(\bar u\gamma u + \bar v\gamma v\right) \, . 
\end{equation}
It then follows from the $p$-equation of (\ref{xpeqs}) that
\begin{equation}
p = \frac{m}{\bar u v} e^{-1}\dot x\, . 
\end{equation}
In the gauge $\bar u v=m$ this reduces to the usual equation $ep=\dot x$. 

We conclude, in agreement with  \cite{Sorokin:1992sy}, that the  action (\ref{SV}) describes a particle of zero spin and mass $m$, provided that $\bar u v\ne0$. 
However,  the spinors $(u,v)$ are auxiliary variables in this action; they are not canonical variables  with the canonical conjugate spinors needed for a
twistor description. One could now add ``kinetic'' terms for $(u,v)$, as was done in  \cite{Sorokin:1992sy} in order to introduce conjugate variables, and spin, but 
this approach differs from the one explored here and it leads to a different end result.

\subsection{The 3D  massive superparticle}

The action for an $N$-extended 3D particle of mass $m$ and zero superspin, but without a Wess-Zumino mass term, is 
\begin{equation}\label{sp1}
S= \int\! dt \left\{ \left(\dot x + i\bar\theta_a \gamma \dot\theta_a\right)\cdot p - \frac{1}{2}e\left(p^2 +m^2\right) \right\}\, , \qquad (a=1,\dots,N)
\end{equation}
where $\theta_a$ are now $N$ anticomuting 3D Majorana spinors. This action is invariant under the spacetime supersymmetry transformations
\begin{equation}
\delta \theta_a= \epsilon_a\, , \qquad \delta x = -i\bar\epsilon_a\gamma \,\theta_a\, ,
\end{equation}
and the corresponding Noether charges are
\begin{equation}\label{susycharges}
Q_a = -\slashchar{p}\, \theta_a\, . 
\end{equation}

We may solve the mass-shell constraint as in (\ref{particlep}). Then,  introducing the new anticommuting variables
\begin{equation}
\mu_a= \,\bar u \theta_a \, , \qquad \nu_a = \,\bar v\theta_a\, , 
\end{equation}
and using the identities
\begin{equation}\label{umuids}
\left(\bar u \gamma u\right)\cdot \gamma \theta_a = 2\mu_au\, , \qquad 
\left(\bar v \gamma v\right)\cdot \gamma \theta_ia= 2\nu_a v\, , 
\end{equation}
we  find that 
\begin{equation}
\left(\dot x + i\bar\theta_a \gamma \dot\theta_a\right)\cdot p = \dot{\bar u} w + \dot{\bar v} z + i\left(\mu_a\dot\mu_a + i\nu_a\dot\nu_a \right) + \frac{d}{dt}\left(\right)\, , 
\end{equation}
where
\begin{equation}
w= \slashchar{x}\, u - i\mu_a\theta_a\, , \qquad z= \slashchar{x} v - i \nu_a\theta_ia\, . 
\end{equation} 
It follows from these definitions that
\begin{equation}
\bar u z - \bar v w - 2i\mu_a\nu_a \equiv 0\, , 
\end{equation}
so this must be imposed as a constraint in the twistor form of the action. In terms of the complex anticommuting scalars 
\begin{equation}
\xi_a =  \left(\mu_a + i\nu_a\right) \, , 
\end{equation}
and the complex spinors $(\rho,\omega)$ defined in (\ref{compspin}), we thus arrive at the action
\begin{equation}\label{newacts}
S= \int \! dt \left\{\dot{\bar \rho} \omega  - \bar\omega \dot\rho    + i \bar\xi_a \dot\xi_a  +i \ell \left(\bar \rho\rho -  im\right) - s\left(\Lambda - \frac{1}{2}  \bar\xi_a\xi_a\right)\right\} \, . 
\end{equation}

Upon quantization, the bilinears $n_a = \bar\xi_a \xi_a$ (no sum on $a$) become fermion occupation numbers taking the values $0$ or $1$. In order to preserve the parity invariance of the classical action, we must include in the quantum spin-shell constraint the fermi zero-point ``energy'', which means that the allowed helicities are 
the eigenvalues of the operator 
\begin{equation}
\frac{1}{2} \sum_{a=1}^N n_a - \frac{N}{4}\, . 
\end{equation}
This gives us a multiplet of helicity states,  with binomial multiplicities, separated by helicity $1/2$, and with maximal helicity $N/2$.  This is the zero superspin supermultiplet of 
$N$-extended 3D supersymmetry.  For example, for $N=1$ we get  a semion supermultiplet of  helicities $(-1/4,1/4)$ \cite{Mezincescu:2010yq}. 

By construction, the action (\ref{newacts}) must have $N$-extended space-time supersymmetry. In fact,  it is invariant under the following supersymmetry transformations with 
$N$ {\it complex} 3D spinor anticommuting parameters $\epsilon_a$:
\begin{eqnarray}
\delta \omega&=& i\epsilon_a \xi_a \, , \qquad \Rightarrow\quad \delta\bar\omega = i \bar\epsilon_a \bar\xi_a\, , \nonumber \\
\delta\xi_a &=& - \bar\epsilon_a \rho  \ \ \qquad \Rightarrow \quad \delta\bar\xi_a = \bar\rho\epsilon_a\, .  
\end{eqnarray}
The corresponding Noether charges are the complex 3D spinors 
\begin{equation}
{\cal S}_a = \bar\xi_a \rho\, . 
\end{equation}
This means, in particular, that the $N=1$ 3D superparticle actually has an $N=2$ supersymmetry, in agreement with \cite{Gorbunov:1996ed}. Similarly, the generic $N=2$ massive superparticle action actually has an $N=4$ supersymmetry.

\subsubsection{$N=2$ with Wess-Zumino mass term}

Now we focus on the $N=2$ case but add the Wess-Zumino mass term
\begin{equation}
L_{WZ}= iq\varepsilon^{ab}\bar\theta_a \dot\theta_b\, . 
\end{equation}
We shall make use of the identity 
\begin{equation}
\varepsilon^{ab}\left[\left(\bar u \dot\theta_a\right)\left(\bar v\theta_b\right) - \left(\bar v\dot\theta_a\right)\left(\bar u \theta_b\right)\right] = - \left(\bar u v\right) \varepsilon^{ab} \theta_a\dot\theta_b
\end{equation}
and the constraint $\bar u v=m$ to rewrite $L_{WZ}$ as 
\begin{equation}
L_{WZ} = \dot{\bar u}\left[ i\frac{q}{m} \varepsilon^{ab} \theta_a\nu_b\right] + \dot{\bar v}\left[-i\frac{q}{m} \varepsilon^{ab} \theta_a\mu_b\right] 
+ \frac{q}{m} i\varepsilon^{ab} \nu_b \dot\mu_a - \frac{q}{m} i\varepsilon^{ab} \mu_b \dot\nu_a\, . 
\end{equation}
Adding this to the $N=2$ case of the action (\ref{newacts}),  we get (after using the constraint $\bar u v =m$ to simplify the spin-shell constraint and dropping a total time derivative
from the Lagrangian)
\begin{equation}\label{newacts2}
S=  \int \! dt \left\{  \dot{\bar u} w +  \dot{\bar v}z + i \bar\xi_a \left(\delta^{ab} + i\frac{q}{m} \varepsilon^{ab}\right)\dot\xi_b -\ell \left(\bar u v -m \right) -s \varphi \right\}\, , 
\end{equation}
where $\xi_a= \mu_a+i\nu_a$, as before, and the spinors canonically conjugate to $(u,v)$ are now
\begin{equation}
w =  \slashchar{x} u -i\mu_a\theta_a -i\frac{q}{m} \varepsilon_{ab} \nu_a\theta_b\, , \qquad \tilde z= \slashchar{x} v-i\nu_a\theta_a + i\frac{q}{m}\varepsilon^{ab}\mu_a\theta_b \, , 
\end{equation}
and the spin-shell constraint function is now
\begin{equation}
\varphi=  \frac{1}{2} \left(\bar u z - \bar v w\right)  - \frac{1}{2} \bar\xi_a \left(\delta^{ab} - i\frac{q}{m} \varepsilon^{ab}\right) \xi_b \, . 
\end{equation}
Replacing $(u,v)$ by the complex combinations $(\rho,\omega)$ defined in (\ref{compspin}), we may rewrite this action as
\begin{equation}
S= \int\! dt \left\{ \dot{\bar\rho}\omega - \bar\omega \dot\rho  + i\bar\xi_a K^{ab} \dot\xi_b  +i\ell \left(\bar \rho\rho -im \right) 
-s\left(\Lambda -  \frac{1}{2} \bar\xi_a K^{ab} \xi_b\right)\right\}\, , 
\end{equation}
where
\begin{equation}
K^{ab} = \delta^{ab} - i\frac{q}{m} \varepsilon^{ab}\, .  
\end{equation}
By comparison with (\ref{newacts}) we see that the effect of the Wess-Zumino term is to insert the matrix $K$ into the terms bilinear in $\xi_a$. 

The action is invariant under the supersymmetry transformations
\begin{eqnarray}
\delta \omega&=& i\epsilon_a K^{ab} \xi_b \, , \qquad \Rightarrow\quad \delta\bar\omega = i \bar\epsilon_a \bar K^{ab} \bar\xi_b\, , \nonumber \\
\delta\xi_a &=& - \bar\epsilon_a \rho\, , \!\!\qquad \Rightarrow \quad \delta\bar\xi_a = \bar\rho\epsilon_a\, , 
\end{eqnarray}
for complex spinor parameters $\epsilon_a$, and the corresponding complex spinor Noether charges are
\begin{equation}
{\cal S}_a = \bar K^{ab} \bar\xi_b \rho\, . 
\end{equation}
As long as the matrix $K$ is invertible, the $N=2$ massive particle still  has $N=4$ supersymmetry. 

The matrix $K$ fails to be invertible only if $m^2=q^2$, i.e. when the BPS bound is saturated. Without loss of generality we may choose $q$ to be positive, so that the BPS $N=2$ superparticle has $m=q$. The standard action in this case  has  a  ``kappa-symmetry'' that allows half of the components of the anticommuting spinor variables $\theta_a$  to 
be ``gauged away''.  The twistor form (\ref{newacts})  of the generic $N=2$ massive superparticle action  has a similar, but simpler, fermionic gauge invariance when $m=q$
because the non-invertibility of $K$ then implies that it  depends on the two complex anticommuting 
variables $(\xi_1,\xi_2)$ only through the linear combination 
\begin{equation}
\xi= \xi_1 - i\xi_2\, . 
\end{equation}
Specifically, the action is 
\begin{equation}
S= \int\! dt \left\{ \dot{\bar\rho}\omega - \bar\omega \dot\rho  + i\bar\xi\dot\xi  +i\ell\left(\bar \rho\rho -im \right) 
-s\left(\Lambda -  \frac{1}{2} \bar\xi\xi\right)\right\}\, .
\end{equation}
This is identical to the $N=1$ massive superparticle action (\ref{newacts}), confirming the equivalence noted in \cite{Gorbunov:1996ed}.

\section{Spinning particle}\label{sec:4D}

The equations of motion of a free massless 4D particle of  spin-$N/2$  in Minkowski spacetime can be derived from the phase-space action \cite{Volkov:1989ky,Howe:1988ft}
\begin{equation}
S= \int \! dt \left\{ \dot X \cdot P + \frac{i}{2} \lambda_a \cdot\dot \lambda_a - \frac{1}{2} e P^2 - i \varsigma_a \lambda_a \cdot P - \frac{i}{2} f_{ab} \lambda_a\cdot\lambda_b\right\} \, , 
\end{equation}
where $\{\lambda_a ; a=1,\dots,N\}$ is a set of anticommuting 3-vector variables. The Lagrange multipliers $(e,\varsigma_a, f_{ab})$ can be viewed as $N$-extended worldline supergravity gauge fields that ensure reparametrization invariance, local invariance under $N$ worldline supersymmetries and local $SO(N)$ invariance.  We will need the infinitesimal gauge transformations of the canonical variables. The non-zero transformations are
\begin{equation}
\delta X = \alpha P  + i \epsilon_a \lambda_a \, , \qquad \delta\lambda_a = - \epsilon_a P + \beta_{ab} \lambda_b\, , 
\end{equation}
where $\alpha$ and $\beta_{ab}$ ($=-\beta_{ba}$) are commuting parameters and the $\epsilon_a$ are anticommuting parameters. For invariance of the action the Lagrange multipliers must transform as follows:
\begin{equation}
\delta e = \dot\alpha - 2i\epsilon_a\varsigma_a\, , \qquad \delta \varsigma_a = \dot\epsilon_a + \beta_{ab} \varsigma_b \, , \qquad \delta f_{ab} = \dot\beta_{ab}  -\beta_{ca}f_{cb} + \beta_{cb}f_{ca}\, . 
\end{equation}
The Lagrange multiplier $f_{ab}$ is antisymmetric in its  $N$-vector indices, and is therefore identically zero for $N=1$; in this case the action reduces to the standard ``spinning particle'' action describing, in the quantum theory, a particle of spin $1/2$.  More generally,  quantization yields free field equations for a massless particle  of spin $N/2$. 

We now show how the supertwistor formulation of this model can be found,  following the construction for $N=1$ presented in \cite{Townsend:1991sj}. 
By means of the identity (\ref{4Did})  we may solve the  mass-shell and supersymmetry constraints by setting 
\begin{equation}\label{solvecons}
P= - \frac{1}{2}\bar U \Gamma U\, , \qquad \lambda_a = \bar U \Gamma \chi_a\, , 
\end{equation}
where $U$ is an arbitrary 4D Majorana spinor,  $\chi_a$ are $N$ anticommuting variables, and the components of the 4-vector $\Gamma$ are the four 4D Dirac matrices. 
The gauge transformations of the new anticommuting variables are
\begin{equation}
\delta\chi_a = \frac{1}{2} U \epsilon_a + \beta_{ab} \chi_b\, . 
\end{equation}
We are still left with the $SO(N)$ constraint. Using the identity
\begin{equation}\label{4Did2}
\left(\bar U \Gamma \chi\right) \cdot \left(\bar U \Gamma \psi\right) \equiv  \left(\bar U \chi\right) \left(\bar U \psi\right) + \left(\bar U\gamma_5 \chi\right) \left(\bar U \gamma_5\psi\right)\, , 
\end{equation}
which is  valid for arbitrary {\it anticommuting} Majorana spinors $(\chi,\psi)$, and defining the new anticommuting variables
\begin{equation}
\mu_a = \frac{1}{\sqrt{2}}\bar U \chi_a\, , \qquad \nu_a= \frac{1}{\sqrt{2}}\bar U \gamma_5 \chi_a\, , 
\end{equation}
we find that 
\begin{equation}
\frac{1}{2} \lambda_a \cdot \lambda_b = \mu_a \mu_b + \nu_a\nu_b\, .
\end{equation}
Observe that the gauge transformations of $(\mu_a,\nu_b)$ are
\begin{equation}
\delta\mu_a = \beta_{ab} \mu_b\, , \qquad \delta\nu_a = \beta_{ab} \nu_b\, , 
\end{equation}
In other words, the anticommuting variables $(\mu_a,\nu_a)$ are gauge invariant except for their transformation as $N$-vectors with respect to the local $SO(N)$.

Now we use (\ref{solvecons}) and the identity (\ref{4Did2}) to show that 
\begin{equation}
\dot X \cdot P + \frac{i}{2} \lambda_a\cdot\dot\lambda_a = \dot {\bar U} W + i\left[ \mu_a \dot\mu_a + \nu_a\dot{\nu}_a\right]\, , 
\end{equation}
where
\begin{equation}
W= \slashchar{X} U + \frac{i}{2} \left(\bar U \Gamma \chi_a\right) \cdot\Gamma \chi_a - \frac{i}{2}\left( \bar U \chi_a\right)\chi_a - \frac{i}{2}\left( \bar U\gamma_5 \chi_a\right)\gamma_5\chi_a \, . 
\end{equation}
It follows from this expression that
\begin{equation}
\bar U \gamma_5 W + 4i \mu_a\nu_a \equiv 0\, ,
\end{equation}
and this must be added as a constraint if we wish to promote $W$ to an independent variable. Taking this and the $SO(N)$ constraint into account, and defining 
\begin{equation}
\xi_a= \left( \mu_a - i \nu_a \right)\, , 
\end{equation}
we arrive at the action 
\begin{equation}\label{twistact}
S= \int \! dt \left\{ \dot {\bar U} W + i\bar\xi_a\dot\xi_a - s \left(\Lambda - \bar\xi_a\xi_a  \right)  - \frac{i}{2} f_{ab} \bar\xi_{a} \xi_{b}\right\}\, , 
\end{equation}
where, as always,  $2\Lambda = \bar U \gamma_5 W$. 
In the quantum theory, $(\bar\xi_a,\xi_a)$ become $N$ pairs of fermi oscillator creation and annihilation operators, and the products $n_a= \bar\xi_a\xi_a$ (no sum over $a$) 
become  $N$ fermion  number operators, with eigenvalues $0$ and $1$.  There is an operator ordering ambiguity, which we resolve  so as to preserve the parity invariance of the classical action; this requires the inclusion of the standard zero point energy term for each fermi oscillator. The net result is that  the allowed helicities are the eigenvalues of the operator
\begin{equation}\label{spinningspec}
 \sum_{a=1}^N n_a - \frac{N}{2}\, . 
\end{equation}
For $N=1$ this gives us two states with helicities $\pm 1/2$, as expected for a massless particle of spin  $1/2$.  For $N>1$ we we have to take into account the $SO(N)$  constraint; this is equivalent to the condition that  
\begin{equation}\label{spinningspec2}
n_1= \dots = n_N\, , 
\end{equation}
which projects out all but the two states of helicities $\pm N/2$. As expected, the action describes a massless particle of spin $N/2$. 

\subsection{Comparison with massless 4D superparticle}

The supertwistor form of the action for a massless $N$-extended 4D superparticle has been known for a long time \cite{Ferber:1977qx,Shirafuji:1983zd}. In the notation used here 
this action is 
\begin{equation}\label{twistact2}
S= \int \! dt \left\{ \dot {\bar U} W + i\bar\xi_a\dot\xi_a- s\left(\Lambda - \frac{1}{2}\bar\xi_a\xi_a  \right)  \right\}\, . 
\end{equation}
As a check, one may easily verify that reduction of this to 3D in the way described earlier yields the 3D massive superparticle action (\ref{newacts}) . 

Comparing (\ref{twistact2}) with the $N$-extended spinning particle action (\ref{twistact}) we see (i) that the $SO(N)$ constraint is absent, so the fermion occupation numbers 
$n_a = \bar\xi_a\xi_a$ (no sum on $a$) are all independent, and (ii) that these numbers appear in the spin-shell constraint with an additional factor of $1/2$. 
This means that for even $N$ we get an $N$-extended CPT self-dual supermultiplet with helicities
\begin{equation}\label{formhel}
\frac{1}{2} \sum_{a=1}^N n_a - \frac{N}{4}\ . 
\end{equation}
For example, for $N=4$ we get the $N=4$ CPT self-dual Maxwell supermultiplet with maximum spin $1$. 

For odd $N$ the formula (\ref{formhel}) includes states of helicity $\pm 1/4$, which is incompatible with the half-integral quantization condition 
on 4D helicity. For $N=1$, for example, it gives a supermultiplet of helicities $\pm 1/4$. What this means is that we must choose a different operator 
ordering such that all helicities are shifted so as to make them half-integral. For $N=1$ we could choose to shift the helicities so as to get massless 
states of helicity $(0,1/2)$. However, we now have a CPT anomaly. This is resolved in the field theory context by including another $N=1$ massless 
supermultiplet with helicities $(-1/2,0)$ but CPT invariance is not automatically incorporated at the level of particle mechanics. 

Notice that the quantization condition on 4D helicity does not arise directly from quantization of the massless 4D particle action. There is no obvious reason
why we could not shift the spin-shell constraint function by an arbitrary constant, as we can do in 3D, because this would not break any symmetries or gauge 
invariances of the classical action. This suggests that it might be possible for massless 4D particles to have fractional spin, but this possibility has been considered and excluded
\cite{Volkov:1989qa,Klishevich:2001gy}. One reason for the quantization condition on 4D helicity is that the universal cover of the $SO(2)$ ``little group'' for 
massless particles is not a subgroup of the universal cover of the 4D Lorentz group.

\subsection{The 3D massive spinning particle}

By imposing the condition $P_3=m$ in the massless 4D spinning action, following the steps  explained  in 
section \ref{sec:3D}, we get the following action for a spinning particle of mass $m$ in 3D:
\begin{equation}
S= \int \! dt \left\{\dot{\bar u} w + \dot{\bar v}  z +i \bar\xi_a \dot\xi_a -\ell\left(\bar u v- m\right)  -s \left(\Lambda - 
\bar\xi_a\xi_a \right)  - \frac{i}{2} f_{ab} \bar\xi_{[a} \xi_{b]}\right\}\, . 
\end{equation}
Here $\Lambda$ is now the 3D helicity, so $2\Lambda = \bar u z - \bar v w$. This action is invariant under a $\bZ_2$ symmetry corresponding to 3D parity; this would be broken by the 
inclusion of a LWZ term but we will not consider that possibility in this context.  If we use 3D parity to resolve ordering ambiguities in the quantum theory then we find, 
that the 3D helicities described by this action are the eigenvalues of exactly the same operator (\ref{spinningspec}) as the 4D massless case,  and that the  fermion occupation numbers are subject to exactly the same constraints (\ref{spinningspec2}).  As in the massless 4D case, this means that  the action describes a particle with two polarization states of 
(3D) helicities $\pm N/2$, which now form a parity doublet.

\subsection{$N=2$ and the worldline Chern-Simons  term} 

The $N=2$ case of the spinning particle action is special because then $f_{ab}= \varepsilon_{ab} f$, for scalar Lagrange multiplier $f$, and we may add to the action a term linear in $f$.
This may be viewed as a ``worldline Chern-Simons'' term since $f$ is an $SO(2)$ worldline gauge potential \cite{Howe:1988ft}.  In the context of the twistor form of the action for a 4D  massless spinning particle, this leads to the modified action
\begin{equation}
S= \int \! dt \left\{\dot{\bar U} W  + i\bar\xi_a \dot\xi_a  -s\left(\Lambda - 
\bar\xi_a\xi_a\right)  - if \left(\bar\xi_1 \xi_2 - \bar\xi_2 \xi_1-  ic\right) \right\}\, , 
\end{equation}
where $c$ is a constant, the coefficient of the worldline Chern-Simons term. 

In the quantum theory, the spin-shell constraint tells us that 
\begin{equation}
\Lambda = n_1+n_2 - 1\, , 
\end{equation}
where we again include the fermion ``zero point energy'' contribution, since this is required to preserve parity. The constraint imposed by $f$ becomes, in the quantum theory, the condition that physical states be annihilated by  the operator $\bar\xi_1 \xi_2 - \bar\xi_2 \xi_1-  ic$. Using the  identity 
\begin{equation}
\left(\bar\xi_1 \xi_2 - \bar\xi_2 \xi_1\right)^2 \equiv -\left(n_1-n_2\right)^2\, , 
\end{equation}
we see that this physical state condition requires that 
\begin{equation}
c^2 = \left(n_1-n_2\right)^2\, , 
\end{equation}
which implies that {\it either} $c=0$ {\it or} $c=\pm1$. In other words, if we add the worldline Chern-Simons term then quantum consistency requires that its coefficient is $\pm1$. 
Without this term we have $n_1=n_2$ so there are two states, of  helicities $\pm1$, as expected for a massless spin-$1$ particle.   If $c=\pm1$ then $n_1\ne n_2$, so $(n_1,n_2)$ is either $(1,0)$ or $(0,1)$, and in either case there is a single state of helicity zero. We thus find, in agreement with \cite{Howe:1988ft}, that the 4D $N=2$ spinning particle action with worldline Chern-Simons term describes a zero-spin  particle.

Essentially the same result applies to the 3D massive case. Without the worldline Chern-Simons term the action describes a massive spin-$1$ particle, with a parity doublet of 
helicities $\pm1$, and with the worldline Chern-Simons term it describes a massive spinless particle.

\section{Massive 4D (super)particle}
\setcounter{equation}{0}

The action for a 4D spin zero particle of mass $m$  is 
\begin{equation}
S= \int\! dt \left\{ \dot X\cdot P - \frac{1}{2}e \left(P^2 +m^2\right) \right\}\, . 
\end{equation}
The mass-shell  constraint function generates time reparametrizations, so the physical phase-space has dimension $8-2\times 1 =6$.

Now we solve the mass-shell constraint by setting 
\begin{equation}\label{solve4P}
P= -\frac{1}{2}\left(\bar U \Gamma U + \bar V\Gamma V\right)\, , 
\end{equation}
where $(U,V)$ are two Majorana spinors, and by imposing the  constraint
\begin{equation}\label{onecon}
\left(\bar U V\right)^2 + \left(\bar U \gamma_5 V\right)^2 = m^2 \, . 
\end{equation}
This solution works because of the identity
\begin{equation}
\left(\bar U \Gamma U\right)\cdot\left(\bar V\Gamma V\right) \equiv -2\left[\left(\bar U V\right)^2 + \left(\bar U \gamma_5 V\right)^2\right]\, . 
\end{equation}

Using the expression for $P$ we now find that 
\begin{equation}\label{WZdef}
\dot X \cdot P = \dot{\bar U} W + \dot{\bar V} Z + \frac{d}{dt}\left(\right) \, , \qquad W= \slashchar{X} U \, , \quad Z= \slashchar{X} V\, . 
\end{equation}
We already see from this result that the non-zero Poisson brackets of the canonical variables will be
\begin{equation}\label{4Dcanon}
\left\{\bar U_\alpha, W^\beta\right\}_{PB}= \delta_\alpha^\beta \, ,  \qquad 
\left\{\bar V_\alpha, Z^\beta\right\}_{PB} = \delta_\alpha^\beta  \, . 
\end{equation}
From the definitions of $W$ and $Z$ in (\ref{WZdef}) we may deduce the identities
\begin{eqnarray}\label{defphis}
\varphi_1 &:=& \frac{1}{2} \left(\bar V\gamma_5 W + \bar U\gamma_5 Z\right) \equiv 0\, , \nonumber \\
\varphi_2 &:=& \frac{1}{2}\left( \bar U\gamma_5 W - \bar V\gamma_5 Z\right) \equiv 0 \nonumber\\
\varphi_3  &:=&  \frac{1}{2} \left(\bar V W - \bar U Z\right)  \equiv 0\, ,  
\end{eqnarray}
and
\begin{equation}
\chi := \frac{1}{2}\left( \bar U\gamma_5 W + \bar V\gamma_5 Z\right) \equiv 0 \, , 
\end{equation}
and these become additional constraints when $(W,Z)$ are considered as independent variables. Using (\ref{4Dcanon}) we find that 
\begin{equation}\label{SU2}
\left\{\varphi_I,\varphi_J\right\}_{PB}= \epsilon_{IJK}\, \varphi_K\, , \qquad (I,J,K=1,2,3)
\end{equation}
which is the algebra of $SU(2)$.  From (\ref{6Dto4D}) one sees that the $\varphi_I$  are just the reduction from 6D of the 
triplet of constraint functions that generate the $SU(2)$ gauge transformations of  the twistor form of the massless 6D action, so we 
should expect these constraint functions  to remain first-class in 4D. 
There are now two ways to proceed, according to how we implement the condition (\ref{onecon}).
\begin{enumerate}

\item We can view (\ref{onecon}) as a Hamiltonian constraint, with quartic constraint function
\begin{equation}
\zeta = \frac{1}{2}\left(\bar U V\right)^2 + \left(\bar U \gamma_5 V\right)^2 - m^2 
\end{equation}
The action is then
\begin{equation}\label{allfirst}
S= \int\! dt\left\{ \dot{\bar U}W + \dot{\bar V} Z - s^I \varphi_I - \varsigma\chi - \rho\zeta\right\}\, , 
\end{equation}
where $(s^I,\varsigma, \rho)$ are  Lagrange multipliers for a total of five constraints. All constraints are first-class, and so all generate gauge transformations.
In particular, $\chi$ generates a chiral $U(1)$ gauge transformation 
on $U+iV$ and $Z+iW$, while $\zeta$ generates the gauge transformation
\begin{eqnarray}\label{obscuregt}
\delta W &=& \left(\bar U V\right) V + \left(\bar U\gamma_5 V\right) \gamma_5 V\, , \nonumber \\
\delta Z &=& - \left(\bar U V\right) U - \left(\bar U \gamma_5 V\right) \gamma_5 U\, . 
\end{eqnarray} 
The physical phase-space has 
dimension $16 - 2\times 5=6$, as expected.

\item We can also satisfy (\ref{onecon})  by setting $\bar U V=m$ and $\bar U \gamma_5 V=0$; in this case all  Hamiltonian constraints are quadratic. 
Defining the new constraint  functions
\begin{equation}\label{twocons}
\chi' = \bar U \gamma_5 V\, , \qquad \psi = \bar U V -m\, , 
\end{equation}
we can write the action as 
\begin{equation}\label{partialgf} 
S= \int\! dt\left\{ \dot{\bar U}W + \dot{\bar V} Z - s^I \varphi_I - \varsigma\chi - \varsigma' \chi' - \ell\psi\right\}\, , 
\end{equation}
where $(s^I,\varsigma,\varsigma', \ell)$ are  Lagrange multipliers.  

The non-zero Poisson brackets of constraint functions are now those of (\ref{SU2}) and
\begin{equation}
\left\{\chi,\chi' \right\}_{PB}= \bar U V = m\, , \quad \left\{\chi,\psi\right\}_{PB} = -\chi' = 0\, , 
\end{equation}
where, in each case, the second equality uses the constraints (\ref{twocons}). 
This shows that $\psi$ is first class, it generates time reparametrizations, but the two constraint functions $(\chi,\chi')$ are second-class.  We now have a phase space of dimension 16 (four real 4-component spinors) subject to 4 first-class constraints and 2 second-class constraints, so the physical phase-space dimension is
$16 - 2\times 4 -2 =6$, as before. 

\end{enumerate}
The action (\ref{partialgf})  can be viewed as a version of the action (\ref{allfirst})  in which the $U(1)$ chiral gauge invariance  has been fixed.

\subsection{Poincar\'e Casimirs and the spin-shell constraints}

We have explained in subsection \ref{subsec:genhel} the significance of the spin-shell constraints for a massless 6D particle. Now we address the same issue for 4D massive particles. 
In this case, the unitary irreducible representations of the Poincar\'e group are classified by the values of the Poincar\'e Casimirs. One is ${\cal P}^2=-m^2$, where $m$ is the
particle's mass. The other is  the square of the Pauli-Lubanski spin-vector $\Sigma$.  In 4D twistor variables we have
\begin{equation}\label{PLtwist}
\Sigma^n =  -\frac{1}{2} \left(\bar U \Gamma_m U + \bar V\Gamma_m V\right)\ \left(\bar U \Gamma^{mn}\gamma_5 W + \bar V \Gamma^{mn}\gamma_5 Z\right)\, .
\end{equation}
Using $\Gamma^{mn}= \Gamma^m\Gamma^n - \eta^{mn}$, the identities (\ref{4Did}), and the further identities
\begin{eqnarray}\label{id2}
\left(\bar U \Gamma U\right) \cdot \left(\bar V \Gamma\psi\right)  &\equiv & -2 \left(\bar U V\right)\left(\bar U \psi\right) 
-2\left(\bar U\gamma_5 V\right) \left(\bar U \gamma_5 \psi\right)\, ,  \nonumber \\
\left(\bar V \Gamma V\right) \cdot \left(\bar U \Gamma\psi\right)  &\equiv & 2 \left(\bar U V\right)\left(\bar V \psi\right) 
+ 2\left(\bar U\gamma_5 V\right) \left(\bar V \gamma_5 \psi\right)\,  , 
\end{eqnarray}
we can rewrite (\ref{PLtwist}) as 
\begin{eqnarray}
\Sigma^n &=& \frac{1}{2} \left(\bar U\Gamma^n U + \bar V\Gamma^n V\right) \left(\bar U\gamma_5 W +\bar V\gamma_5 Z\right) \nonumber \\
&& \!\! +\, \left(\bar U V\right)\left(\bar U \Gamma^n\gamma_5 Z -\bar V\Gamma^n\gamma_5 W\right) + 
\left(\bar U \gamma_5 V\right) \left(\bar U\Gamma^n Z -\bar V\Gamma^n W\right)\, . 
\end{eqnarray}
By further use of the identities (\ref{id2}), an additional Fierz rearrangement, and use of the constraint (\ref{onecon}), we find that 
\begin{eqnarray}
\Sigma^2/m^2 &=&\left(\bar U\gamma_5 W + V\gamma_5 Z\right)^2 + \left(\bar U Z\right)^2 + \left(\bar U\gamma_5 Z\right)^2 + \left(\bar V W\right)^2 + \left(\bar V\gamma_5 W\right)^2 \nonumber \\
&& -2\left(\bar U\Gamma\gamma_5 Z\right)\cdot\left(\bar V\Gamma\gamma_5 W\right) - 
4\left(\bar U \gamma_5 V\right)\left(\bar W\gamma_5 Z\right)\, . 
\end{eqnarray}
By further Fierz rearrangements, this can be put into the form
\begin{equation}
\Sigma^2 = 4m^2\left[ \varphi_1^2 + \varphi_2^2 + \varphi_3^2\right] \, , 
\end{equation}
where $\varphi_I$ are the functions defined in (\ref{defphis}). This shows that 
the quadratic Casimir of the internal $SU(2)$ gauge group of the twistor action
is proportional to the Poincar\'e Casimir obtained as the norm squared of the Pauli-Lubanski spin-vector.  The spin-shell constraints imply that this is  zero so, as expected, 
the particle has zero spin.

\subsection{$N=1$ massive 4D superparticle}

The $N=1$ massive superparticle action is 
\begin{equation}\label{N=1m}
S= \int\! dt \left\{ \dot X \cdot P + i \bar\Theta \slashchar{P}\, \dot \Theta - \frac{1}{2}e \left(P^2 +m^2\right) \right\}\, , 
\end{equation}
where $\Theta$ is an anticommuting Majorana spinor.  It has a manifest $N=1$ supersymmetry with Majorana spinor Noether charge 
\begin{equation}\label{4Dsusycharge}
{\cal Q} = -\slashchar{P}\, \Theta\, . 
\end{equation}
The physical phase-space has dimension $(6|4)$ because $\Theta$ has four real anticommuting components.

We may solve the mass-shell constraint as before. Then we use identity
\begin{equation}
\left(\bar\Theta \Gamma \dot\Theta\right)\cdot \left(\bar U \gamma  U\right) = -2\left[ \left(\bar U \Theta\right)\left(\bar U \dot\Theta\right) 
+ \left(\bar U \gamma_5\Theta\right)\left(\bar U \gamma_5\dot\Theta\right)\right] \, , 
\end{equation}
and the same identity with $U\mapsto V$, to deduce that 
\begin{equation}
i \bar\theta \slashchar{P}\, \dot \Theta = i \left(\bar U \Theta\right)\left(\bar U \dot\Theta\right) +i \left(\bar U\gamma_5\Theta\right)\left(\bar U \gamma_5\dot\Theta\right) +
i \left(\bar V \Theta\right)\left(\bar V \dot\Theta\right) +i \left(\bar V\gamma_5\Theta\right)\left(\bar V \gamma_5\dot\Theta\right) \, . 
\end{equation}
Now we define the anticommuting (pseudo)scalars
\begin{equation}
\mu = \bar U \Theta \, , \quad \tilde\mu = \bar U \gamma_5 \dot\Theta\, , \qquad \nu = \bar V\Theta\, , \quad \tilde\nu = \bar V\gamma_5\Theta\, . 
\end{equation}
This gives us 
\begin{equation}
\dot X \cdot P + i \bar\Theta \slashchar{P}\, \dot \Theta = \dot{\bar U} W + \dot{\bar V} Z + i\left(\mu\dot\mu + \tilde\mu \dot{\tilde\mu} + \nu\dot\nu + \tilde\nu\dot{\tilde\nu}\right) \, , 
\end{equation}
where, now, 
\begin{equation}
W= \slashchar{X} U -i\mu\Theta -i\tilde\mu \gamma_5\Theta \, , \qquad Z= \slashchar{X} V -i\nu\Theta -i\tilde\nu\gamma_5\Theta\, . 
\end{equation}
From these definitions we get the constraints $\varphi_I=0$, for $I=1,2,3$, and $\chi=0$, where
\begin{eqnarray}
\varphi_1 &=& \frac{1}{2}\left(\bar V \gamma_5 W+ \bar U \gamma_5 Z\right) + i\mu\tilde\nu -i\tilde\mu\nu\, , \nonumber \\
\varphi_2 &=& \frac{1}{2}\left( \bar U\gamma_5 W - \bar V\gamma_5 Z\right) +i\mu\tilde\mu -i\nu\tilde\nu\, , \nonumber \\
\varphi_3 &=& \frac{1}{2} \left( \bar V W - \bar U Z\right) + i\mu\nu +i\tilde\mu\tilde\nu\, , 
\end{eqnarray}
and 
\begin{equation}
\chi= \frac{1}{2}\left( \bar U\gamma_5 W + \bar V\gamma_5 Z\right) +i\mu\tilde\mu +i\nu\tilde\nu\, . 
\end{equation}

These constraints are in addition to either (i) the one constraint (\ref{onecon}) or (ii)  the two constraints with constraint functions $(\chi',\psi)$ of (\ref{twocons}); here we opt for the latter
because it simplifies our later discussion of the Wess-Zumino mass term for the $N=2$ massive superparticle.  Then, taking into account all constraints, and introducing the complex anticommuting variables
\begin{equation}
\xi = \mu + i\nu \, , \qquad \tilde\xi = \tilde\mu + i\tilde\nu\, , 
\end{equation}
we find that the $N=1$ massive superparticle action in supertwistor form is 
\begin{equation}\label{supertwist4D1}
S= \int\! dt \left\{ \dot{\bar U} W + \dot{\bar V} Z + i\bar\xi\dot\xi + i\bar{\tilde\xi}\dot{\tilde\xi}   - s^I\varphi_I - \varsigma\chi - \varsigma'\chi' - \ell \psi\right\}\, , 
\end{equation}
where the constraint functions are now
\begin{eqnarray}
\varphi_1 &=& \frac{1}{2}\left(\bar V \gamma_5 W + \bar U \gamma_5 Z\right) + \frac{1}{2}\xi\tilde\xi - \frac{1}{2}\bar\xi\bar{\tilde\xi}\, ,\nonumber \\
\varphi_2 &=& \frac{1}{2}\left( \bar U\gamma_5 W - \bar V\gamma_5 Z\right) +\frac{i}{2} \xi\tilde\xi + \frac{i}{2}\bar\xi\bar{\tilde\xi}\, , \nonumber \\
\varphi_3 &=& \frac{1}{2} \left( \bar V W - \bar U Z\right) + \frac{1}{2} \bar\xi \xi + \frac{1}{2}\bar{\tilde\xi}\tilde\xi\, ,
\end{eqnarray}
and 
\begin{equation}
\chi = \frac{1}{2}\left( \bar U\gamma_5 W + \bar V\gamma_5 Z\right) +\frac{i}{2} \xi\bar{\tilde\xi} + \frac{i}{2}\bar\xi\tilde\xi\, , 
\end{equation}
and 
\begin{equation}
\chi' = \bar U \gamma_5 V\, , \qquad \psi= \bar U V -m\, . 
\end{equation}
The non-zero Poisson brackets of canonical variables are exactly as before. We have 6 first class  constraints, two of which are second-class,  so the physical phase space dimension is 
$(6|4)$, as expected.

\subsubsection{Hidden supersymmetries}

To express the  supersymmetry Noether charge ${\cal Q}= - \slashchar{P}\, \Theta$ in terms of twistor variables, we use the identity
\begin{equation}
\frac{1}{2}\left(\bar U \Gamma U \right) \cdot \Gamma\Theta \equiv  \left(\bar U \Theta_a\right) U + \left( U\gamma_5 \Theta\right) \gamma_5 U\, , 
\end{equation}
and the same identity with $U\mapsto V$, to find that 
\begin{equation}
{\cal Q}= \bar\xi \rho + \bar{\tilde\xi} \gamma_5\rho + \xi \rho^*  + \tilde\xi \gamma_5\rho^*\, , 
\end{equation}
where
\begin{equation}
\rho= U+iV\, . 
\end{equation}
We can write this as ${\cal Q}= {\cal S}+ {\cal S}^*$, where 
\begin{equation}
{\cal S}= \bar\xi \rho + \bar{\tilde\xi} \gamma_5\rho\, , 
\end{equation}
but the complex Dirac spinor ${\cal S}$ satisfies $\dot {\cal S}=0$ as a consequence of the equations of motion, so its imaginary part is another Majorana spinor charge. In fact
${\cal S}$ generates a symmetry of the action (\ref{supertwist4D1})  with a {\it complex} anticommuting spinor parameter. In other words, the $N=1$ massive superparticle actually has
$N=2$ supersymmetry, exactly as in the 3D case. This extra supersymmetry becomes manifest in the supertwistor form of the action. 

\subsubsection{Quantum Theory}

Let us denote by $S_i$ the spin part of the constraint functions $\varphi_i$. For $S_1$ and $S_2$ there is no ordering ambiguity when we pass to the quantum theory, so
\begin{equation}
S_1= \frac{1}{2}\xi\tilde\xi - \frac{1}{2}\bar\xi\bar{\tilde\xi}\, \qquad S_2= \frac{i}{2} \xi\tilde\xi + \frac{i}{2}\bar\xi\bar{\tilde\xi}\, .
\end{equation}
Using the canonical anticommutation relations
\begin{equation}
\left\{ \xi,\bar\xi\right\} =1 \, , \qquad \left\{ \xi,\bar\xi\right\} =1\, , 
\end{equation}
we compute that 
\begin{equation}
\left[S_1,S_2\right] = \frac{i}{2} \left(\bar\xi\xi + \bar{\tilde\xi}\tilde\xi  -1\right)\, \qquad  n= \bar\xi\xi\, , \quad \tilde n= \bar{\tilde\xi}\tilde\xi\, . 
\end{equation} 
This resolves a potential ordering ambiguity in the quantum operator representing $S_3$; we see that it must be
\begin{equation}
S_3= \frac{1}{2} \left(n+\tilde n -1\right),  \qquad n= \bar\xi\xi\, , \quad \tilde n= \bar{\tilde\xi}\tilde\xi\, . 
\end{equation}
In other words, we must include the usual fermion zero-point ``energy'' for the fermi oscillators. The possible values of $S_3$ are therefore 
$\left(\frac{1}{2},0,0,\frac{1}{2}\right)$, which are just the helicities of the superspin zero $N=1$ supermultiplet.

\subsection{$N$-extended massive 4D superparticle}

The $N=1$ superparticle action (\ref{N=1m}) has the following generalization to one manifestly invariant under an $N$-extended spacetime supersymmetry:
\begin{equation}\label{N=Nm}
S= \int\! dt \left\{ \dot X \cdot P + i \bar\Theta_a \slashchar{P}\, \dot \Theta_a- \frac{1}{2}e \left(P^2 +m^2\right) \right\}\, , 
\end{equation}
where $\Theta_a$ ($a=1,\dots,N$) are $N$ anticommuting Majorana spinors. Proceeding as before we have 
\begin{eqnarray}
i \bar\Theta_a \slashchar{P}\, \dot \Theta_a &=& i \left(\bar U \Theta_a\right)\left(\bar uU\dot\Theta_a \right) 
+i \left(\bar U\gamma_5\Theta_a\right)\left(\bar U \gamma_5\dot\Theta_a\right) \nonumber \\
&&+ i \ \left(\bar V \Theta_a\right)\left(\bar V \dot\Theta_a\right) +i \left(\bar V\gamma_5\Theta_a\right)\left(\bar V \gamma_5\dot\Theta_a\right)  \nonumber \\
&=& i\mu_a \dot\mu_a + i \tilde\mu_a\dot{\tilde\mu}_a + i\nu_a\dot\nu_a + i\tilde\nu_a \dot{\tilde\nu}_a \nonumber \\
&& - \ i  \dot{\bar U} \left(\mu_a\Theta_a +\tilde\mu_a\gamma_5\Theta_a\right)  
-i\dot{\bar V} \left(\nu_a\Theta_a + \tilde\nu_a\gamma_5 \Theta_a\right)\, ,
\end{eqnarray}
where
\begin{equation}
\mu_a = \bar U \Theta_a \, , \quad \tilde\mu_a = \bar U \gamma_5 \dot\Theta_a\, , \qquad \nu_a = \bar V\Theta_a\, , \quad \tilde\nu_a = \bar V\gamma_5\Theta_a\, . 
\end{equation}
After defining the new complex variables
\begin{equation}
\xi_a = \mu_a + i\nu_a\, , \qquad \tilde\xi_a + \tilde\nu_a\, , 
\end{equation}
we arrive at the $N$-extended analog of the $N=1$ action (\ref{supertwist4D1}):
\begin{equation}
S= \int\! dt \left\{ \dot{\bar U} W + \dot{\bar V} Z + i\bar\xi_a\dot\xi_a + i\bar{\tilde\xi}_a\dot{\tilde\xi}_a   - s^I\varphi_I - \varsigma \chi - \varsigma'\chi' - \ell \psi\right\}\, ,
\end{equation}
where the constraint functions are as in the $N=1$ case except that $(\xi,\tilde\xi)$ are replaced by $(\xi_a,\tilde\xi_a)$ and there is a sum over the index $a$. 

Just as the $N=1$ massive superparticle action actually has $N=2$ supersymmetry, so its $N$-extended generalization actually has $2N$-extended supersymmetry, corresponding to the $N$ complex Dirac spinor Noether charges
\begin{equation}
{\cal S}_a=  \bar\xi_a \rho + \bar{\tilde\xi} _a\gamma_5\rho\, . 
\end{equation}

\subsubsection{$N=2$ massive superparticle with WZ mass term}

The generic $N=2$ superparticle action is
\begin{equation}
S= \int\! dt \left\{ \dot X \cdot P + i \bar\Theta_a \slashchar{P}\, \dot \Theta_a  + iq \varepsilon^{ab} \bar\Theta_a \dot \Theta_b - \frac{1}{2}e \left(P^2 +m^2\right) \right\}\, ,
\end{equation}
where the index $a=1,2$ is summed over. 
The new feature is the Wess-Zumino mass term with coefficient $q$ \cite{Azcarraga:1982dw}. Using the identity
\begin{eqnarray}
&& -\left(\bar U V\right) \varepsilon^{ab}\bar\Theta_a\dot\Theta_b + \left(\bar U\gamma_5 V\right) \varepsilon^{ab} \bar\Theta_a\gamma_5 \dot\Theta_b \equiv  \\
&&\varepsilon^{ab} \left[ \left(\bar U\dot\Theta_a\right)\left(\bar V\Theta_b\right) - \left(\bar V\dot\Theta_a\right)\left(\bar U\Theta_b\right) - 
 \left(\bar U\gamma_5\dot\Theta_a\right)\left(\bar V\gamma_5 \Theta_b\right) + \left(\bar V\gamma_5\dot\Theta_a\right)\left(\bar U\gamma_5\Theta_b\right)\right]\, ,
 \nonumber
\end{eqnarray}
and the constraints $\bar U V=m$ and $\bar U \gamma_5 V=0$, we deduce that
\begin{eqnarray}
iq \varepsilon^{ab} \bar\Theta_a \dot\Theta_b &=&   i\frac{q}{m} \varepsilon^{ab} \left[ \mu_a \dot\nu_b - \nu_a \dot\mu_b - \tilde\mu_a \dot{\tilde\nu}_b + \tilde\nu_a \dot{\tilde\mu}_b \right] \nonumber \\
&& - \ i\frac{q}{m} \varepsilon^{ab} \left[ \dot{\bar V}\left(\mu_a \Theta_b - \tilde\mu_a \gamma_5\Theta_b\right) 
- \dot{\bar U}\left(\nu_a \Theta_b - \tilde\nu_a\gamma_5 \Theta_b\right)\right]\, . 
\end{eqnarray}

Putting these results together, and defining
\begin{equation}
K^{ab} = \delta^{ab} - i \frac{q}{m} \varepsilon^{ab}\, , \qquad \bar K^{ab}= \delta^{ab} + i \frac{q}{m} \varepsilon^{ab}\, , 
\end{equation}
we find that
\begin{equation}
\dot X \cdot P + i \bar\Theta_a \left(\slashchar{P}\delta^{ab} + q\varepsilon^{ab}\right)\dot \Theta_b =  \dot{\bar U} W + \dot{\bar V} Z +  i\bar\xi_a K^{ab}\dot\xi_b 
+ i\bar{\tilde\xi}_a \bar K^{ab}\dot{\tilde\xi}_b\, , 
\end{equation}
where
\begin{equation}
\xi_a = \mu_a+i\nu_a\, , \qquad \tilde\xi_a = \tilde\mu_a + i\tilde\nu_a\, , 
\end{equation}
and
\begin{eqnarray}
W &=& \slashchar{X} \, U -i \left[ \left(\mu_a + \tilde\mu_a \gamma_5\right)\Theta_a + 
\frac{q}{m} \varepsilon^{ab} \left(\nu_a - \tilde\nu_a \gamma_5 \right)\Theta_b\right]\, , \nonumber \\
Z &=& \slashchar{X}\, V  -i \left[ \left(\nu_a + \tilde\nu_a \gamma_5\right)\Theta_a - 
\frac{q}{m} \varepsilon^{ab} \left(\mu_a - \tilde\mu_a\gamma_5\right)\Theta_b \right] \, . 
\end{eqnarray}
The twistor form of the $N=2$ action  is therefore 
\begin{equation}
S= \int\! dt \left\{ \dot{\bar U} W+ \dot{\bar V} Z + i \bar\xi_a K^{ab} \dot\xi_b + i\bar{\tilde\xi}_a \bar K^{ab} \dot{\tilde\xi}_b 
 - s^I\varphi_I - \varsigma\chi - \varsigma' \chi' -  \ell \psi\right\}\, ,
\end{equation}
where the Lagrange multipliers $(s^I, \varsigma,\varsigma', \ell)$ impose the constraints with 
constraint functions
\begin{eqnarray}
\varphi_1 &=& \frac{1}{2} \left( \bar V W - \bar U Z\right) + \frac{1}{2} \left(\bar\xi_a K^{ab}\xi_b + \bar{\tilde\xi}_a \bar K^{ab} \tilde\xi_b\right) \, ,  \nonumber \\
\varphi_2 &=& \frac{1}{2}\left(\bar V\gamma_5 W + \bar U \gamma_5 Z\right) + \frac{1}{2} \left(\xi_a\bar K^{ab}\tilde\xi_b - 
\bar\xi_aK^{ab}\bar{\tilde\xi}_b\right)  \, , \nonumber \\
\varphi_3 &=&  \frac{1}{2}\left( \bar U\gamma_5 W - \bar V\gamma_5 Z\right) + \frac{i}{2} \left(\xi_aK^{ab}\tilde\xi_b + \bar\xi_a\bar K^{ab}\bar{\tilde\xi}_b\right) \, , 
\end{eqnarray}
and 
\begin{equation}
\chi = \frac{1}{2}\left( \bar U\gamma_5 W + \bar V\gamma_5 Z\right) +\frac{i}{2}\left( \xi_a\bar K^{ab}\bar{\tilde\xi}_b + \bar\xi_a K^{ab}\tilde\xi_b \right) \, , 
\end{equation}
with $(\chi',\psi)$ as for $N=1$.  The generic $N=2$ massive superparticle actually has two {\it complex} spinor Noether charges
\begin{equation}
{\cal S}_a  = \bar K^{ab} \bar\xi_b \rho + K^{ab} \bar{\tilde\xi}_b \gamma_5 \rho\, , 
\end{equation}
and hence $N=4$ supersymmetry. 

\subsubsection{BPS-saturated case}

Now we specialize to $m=|q|$. Without loss of generality we may assume that $q>0$, so that $q=m$. In this case the action reduces to 
\begin{equation}
S= \int\! dt \left\{ \dot{\bar u} w + \dot{\bar v} z + i \bar\xi\dot\xi  + i\bar{\tilde\xi} \dot{\tilde\xi}
 - s^I\varphi_I - \varsigma\chi - \varsigma'\chi'  - \ell \psi\right\}\, ,
\end{equation}
where 
\begin{equation}
\xi= \xi_1- i\xi_2  \, , \qquad \tilde\xi = \tilde\xi_1 + i\tilde\xi_2\, . 
\end{equation}
The constraints become those of the $N=1$ action, when written in terms of $(\xi,\tilde\xi)$ and their complex conjugates. Therefore, {\it the twistor forms of the $N=2$ BPS superparticle and the masssive $N=1$ superparticle are  identical}. It is only in  this special (BPS-saturated) case that the $N=2$ massive superparticle has only $N=2$ 
supersymmetry rather than $N=4$ supersymmetry. 

\subsubsection{Generic $N$-extended case}

The generic $N$-extended superparticle action involves a Wess-Zumino mass term with an antisymmetric coefficient matrix, which we can skew-diagonalize
to give us $[N/2]$ additional mass parameters $\left\{q_1,\dots,q_{[N/2]}\right\}$, where $[N/2]$ is the integer part  of $N/2$. 

For simplicity we will proceed on the assumption  that $N$ is even; in this case, the action is 
\begin{equation}
S= \int\! dt \left\{ \dot X \cdot P + i \sum_{A=1}^{N/2}\bar\Theta^A_a\left( \slashchar{P}\delta^{ab} + q_A^{ab}\right)\, \dot \Theta^A_b  - \frac{1}{2}e \left(P^2 +m^2\right) \right\}\, . 
\end{equation}
Proceeding as before we arrive at the equivalent supertwistor form of the action
\begin{equation}
S= \int\! dt \left\{ \dot{\bar U} W+ \dot{\bar V} Z + i \sum_{A=1}^{N/2} \left[\bar\xi^A_a K^{ab}_A \dot\xi^A_b + \bar{\tilde\xi}_a{}^A \bar K^{ab}_A \dot{\tilde\xi}_b{}^{\!\! A} \right]
 - s^I\varphi_I- \varsigma\chi - \varsigma'\chi' - \ell \psi\right\},   
\end{equation}
where
\begin{equation}
K_A^{ab}= \delta^{ab} - i \frac{q_A}{m} \varepsilon^{ab}\, , \qquad \bar K_A^{ab}= \delta^{ab} + i \frac{q_A}{m} \varepsilon^{ab}\, , 
\end{equation}
and the expressions for the constraint functions $\varphi_I$ and $\chi$  now involve a similar sum over the index $A$. 
The BPS bound is now
\begin{equation}
m\ge \max{ \left\{q_A; A=1,\dots, N/2\right\}}\, . 
\end{equation}
Provided this bound is not saturated, there are actually $N$ complex spinor Noether supercharges
\begin{equation}
{\cal S}^A_a  = \bar K_A^{ab} \bar\xi^A_b\,  \rho + K_A^{ab} \bar{\tilde\xi}_b{}^{\!\! A} \, \gamma_5 \rho\, , \qquad (a=1,2; \ A=1,\dots, N/2). 
\end{equation}
If the BPS bound {\it is} saturated then there will be $2N-1$ real supercharges,  generically, and only $N$ of them when  $q_A=q$ for all $A$,  
in which  case the action becomes equivalent to the massive $(N/2)$-extended superparticle without a Wess-Zumino mass term.

\section{Conclusions}

An extension to massive particles of the twistor formulation of  the mechanics of massless particles in a Minkowski spacetime requires a pair of twistors  
\cite{Hughston:1981zc}. In the case of three spacetime dimensions (3D) this can be deduced  \cite{Fedoruk:2013sna} by dimensional reduction from Shirafuji's 
twistor formulation of massless particle mechanics in four spacetime dimensions (4D) \cite{Shirafuji:1983zd}. This bi-twistor formulation of massive 3D particle mechanics, 
which we have further developed  here, differs from a number of other  ``twistor-inspired'' formulations (e.g.  \cite{Sorokin:1992sy,Gorbunov:1996ed}) in that the 
the twistor variables replace the usual phase space variables rather than augment them. 

Dimensional reduction of the twistor formulation of six-dimensional (6D) massless particle mechanics \cite{Bengtsson:1987si} was shown in \cite{deAzcarraga:2008ik}
to lead to a  similar bi-twistor formulation of massive 4D particle mechanics. In this case, the mass-shell constraint is replaced by a triplet of spin-shell constraints that generate a local $SU(2)$ invariance. We have shown that the quadratic Casimir of this local $SU(2)$  is the square of the Pauli-Lubanski spin vector, thereby justifying  the ``spin-shell'' terminology for  this case.  In the 6D massless case, we have similarly shown that the triplet of spin-shell constraints functions are 6D analogs of 4D helicity. 

One way of introducing spin in the context of particle mechanics is through the introduction of anticommuting worldline variables. An example is the superparticle, which has
manifest spacetime supersymmetry.  Previous work on the bi-twistor formulation of 4D superparticles was mostly limited to those cases obtainable by dimensional reduction from 
6D,  which we have called  ``BPS superparticles'' since the mass saturates a BPS-type bound implied by the 4D supersymmetry algebra. This construction is not obviously applicable to cases such as the $N=1$ massive superparticle, which we have considered in detail here. Remarkably, this model actually has an $N=2$ spacetime supersymmetry, 
which is manifest in the bi-twistor formulation.  The analogous result in 3D has been known for some time  \cite{Gorbunov:1996ed}, and this too is manifest in the 3D bi-twistor
formalism. 

Not only is it true that the (3D or 4D) $N=1$ massive superparticle actually has $N=2$ supersymmetry but it is also true that it is equivalent to the $N=2$ BPS-superparticle
(and hence does, after all, have a higher dimensional origin). In the 3D case this was already established in  \cite{Gorbunov:1996ed} but by methods specific to that spacetime dimension. The methods used here establish this result for both the 3D and 4D cases. In fact, this equivalence holds  for any dimension, as we have recently shown by other methods \cite{Mezincescu:2014zba}.

The equivalence of the classical actions for the $N=1$ superparticle and the $N=2$ BPS superparticle means that any distinction between them at the quantum level must be due to ambiguities in passing from the classical action to the quantum wave equation\footnote{By classical equivalence we mean here more than just equivalence up to field redefinitions, since we also require that any such field redefinition map the  Poincar\'e charges of one model to the other. If this requirement is relaxed then one can prove, for example, equivalence of the spinning particle and the superparticle for $D=3,4,6$ and $10$ \cite{Sorokin:1988nj,Townsend:1991sj}.}. In the 3D case, quantization of the $N=1$ massive superparticle yields the semion supermultiplet of helicities  $(-1/4,1/4)$  \cite{Mezincescu:2010yq} but quantization of the $N=2$ BPS superparticle yields a supermultiplet with helicities $(-1/4,-1/4, 1/4, 1/4)$, which is a doubled 
version of the $N=1$ semion supermultiplet \cite{Mezincescu:2010gb}. The doubling is due to the fact that the $N=2$ supermultiplet must  carry a central charge, which means that the quantum  wave function must be complex rather than real.  If we want an $N=1$ supermultiplet we choose a real wavefunction but if we want an $N=2$ supermultiplet we must choose a complex wavefunction; the distinction is a purely quantum one. 

Another particle mechanics model in which spin is due to anticommuting variables is the ``spinning particle'', in which the extra variables are $D$-vectors. It was shown in 
\cite{Townsend:1991sj} that the $N=1$ 4D spinning particle has a supertwistor formulation despite not being superconformal invariant. In this formulation, the distinction between the 
$N=1$ spinning particle and the $N=1$ superparticle is just a factor of $2$ in the spin contribution to the spin-shell constraint; this has the effect that the former model describes 
a massless spin $1/2$ particle whereas the latter describes a particle supermultiplet. We have generalised this result to the $N$-extended spinning particle action,  which has
an $SO(N)$ gauge invariance  \cite{Gershun:1979fb,Howe:1988ft,Volkov:1989ky}. In the $N=2$ case,  there is the possibility of including  a ``worldline Chern-Simons''  term, 
and this  leads to an alternative description of spin-zero \cite{Howe:1989vn}; this fact is  particularly transparent in the supertwistor formulation of the $N=2$ spinning particle. 

A general feature of  the (super)twistor formulation of  particle mechanics models in which spin is incorporated through the introduction of anticommuting variables is that 
the spin-shell constraints include fermion number operators associated with fermi oscillators, possibly subject to constraints that relate them.  Different models in the same spacetime dimension and with the same particle mass differ only in the number of fermi oscillators, how they appear in the spin-shell constraint, and the relations (if any) between them.  In the 3D case, it is possible to introduce spin without the need for anticommuting variables via the introduction of a Lorentz-Wess-Zumino term 
(first discussed in \cite{Schonfeld:1980kb} although the terminology used here was introduced in \cite{Mezincescu:2010gb}). This possibility is particularly transparent in the 
(super)twistor formulation; it just amounts to the addition of a constant to the spin-shell constraint. 
 
Finally, we should mention some cases in which we have been unable to find a twistor reformulation. One is the massive 4D spinning particle. It is likely that this is due to a similar
difficulty in the massless 6D case; the problem there is that the solution of the worldline supergravity constraints introduces a commuting spinor and an anticommuting spinor of the same 6D chirality, from which it is not possible to construct anticommuting scalars.  There is a similar difficulty with the $(N,M)$-supersymmetric massless 6D superparticle unless $NM=0$. 
A  twistor formulation of the massive 6D particle is another problematic case; this is presumably a reflection of the difficulties (not necessarily insuperable \cite{Berkovits:1990yc}) that confront a twistor reformulation of the massless 10D particle.

\section*{Acknowledgements} We are grateful to Martin Cederwall for helpful correspondence.   L.M. acknowledges partial support from National Science Foundation Award  PHY-1214521. 
A.J.R. acknowledges support from the UK Science and Technology Facilities Council.





\end{document}